\documentclass[epsfig,useAMS,usenatbib]{mn2e}
\usepackage{epsfig,graphics}

\def\grtsim{\mathrel{\hbox{\rlap{\hbox{\lower2pt\hbox{$\sim$}}}\raise2pt\hbox{$>$}}}}
\def\lesssim{\mathrel{\hbox{\rlap{\hbox{\lower2pt\hbox{$\sim$}}}\raise2pt\hbox{$<$}}}}
\def\oii{[O\,\textsc{ii}]}

\newcommand{\be}{\begin{equation}} 
\newcommand{\ee}{\end{equation}}

\title[Spectroscopic follow-up of a cluster candidate at
$z=1.45$]{Spectroscopic follow-up of a cluster candidate at $z=1.45$}

\author[C. van Breukelen et al.]{Caroline van Breukelen,$^{1}$\thanks{Email: cvb@astro.ox.ac.uk} 
Garret Cotter,$^{1}$ Steve Rawlings,$^{1}$ Tony Readhead,$^{2}$
\newauthor David Bonfield,$^{1}$\thanks{\rm Now at NASA's Goddard Space Flight Center,
Greenbelt, MD 20771, USA.} Lee Clewley,$^{1}$ Rob Ivison,$^{3,4}$ Matt
Jarvis,$^{5}$ 
\newauthor  Chris Simpson,$^{6}$ Mike Watson$^{7}$ \\
$^{1}$Astrophysics, Department of Physics, Keble Road, Oxford, OX1 3RH, UK\\
$^{2}$Owens Valley Radio Observatory, California Institute of
Technology, Pasadena, CA 91125, USA \\
$^3$UK Astronomy Technology Centre, Royal Observatory, Blackford
Hill, Edinburgh EH9 3HJ\\ 
$^4$Scottish Universities Physics Alliance, Institute for
Astronomy, University of Edinburgh, Blackford Hill, Edinburgh EH9 3HJ\\
$^{5}$Centre for Astrophysics, Science \& Technology Research Institute,
University of Hertfordshire, Hatfield, AL10 9AB, UK \\
$^{6}$Astrophysics Research Institute, Liverpool John Moores
University, Twelve Quays House, Egerton Wharf, Birkenhead CH41 1LD, UK\\
$^{7}$X-ray Astronomy Group, Department of Physics and Astronomy,
University of Leicester, Leicester LE1 7RH, UK}
 \date{Released 2007 Xxxxx XX}
\pagerange{\pageref{firstpage}--\pageref{lastpage}} \pubyear{2002}

\def\LaTeX{L\kern-.36em\raise.3ex\hbox{a}\kern-.15em
    T\kern-.1667em\lower.7ex\hbox{E}\kern-.125emX}

\begin{document}
\label{firstpage}
\maketitle
\begin{abstract}
We have obtained deep optical spectroscopic data of the
highest-redshift cluster candidate ($z \sim 1.4$, CVB13) selected by
Van Breukelen et al. (2006) in a photometric optical/infrared
catalogue of the Subaru XMM-Newton Deep Field. The data, which
comprise 104 targeted galaxies, were taken with the DEep Imaging
Multi-Object Spectrograph (DEIMOS) on the Keck 2 telescope and yielded
31 secure redshifts in the range $1.25 < z < 1.54$ within a
$7^{\prime} \times 4^{\prime}$ field centred on CVB13. Instead of one
massive cluster at $z=1.4$, we find evidence for three projected
structures at $z=1.40$, $z=1.45$, and $z=1.48$. The most statistically
robust of these structures, at $z=1.454$, has six spectroscopically
confirmed galaxies. Its total mass is estimated at $\grtsim 10^{14}\rm
M_{\odot}$ and it may therefore be termed a cluster. There
is an X-ray source at the cluster position which is marginally
spatially resolved but whose X-ray spectrum is too hard to be thermal
cluster emission. Its origin could be the summed X-ray emission from
active galaxies in, and projected onto, the cluster. Serendipitously
we have discovered a cluster at $z=1.28$ with a mass of $\grtsim
10^{14}\rm M_{\odot}$ at the same position on the sky, comprising six
spectroscopically confirmed cluster galaxies and at least one
additional radio source. The selection of CVB13 for the cluster
catalogue was evidently aided by the superposition of other,
presumably lower-mass, structures, whereas the single cluster at
$z=1.28$ contained too few galaxies to be isolated by the same
algorithm. Given the complicated nature of such structures, caution
must be employed when measuring the mass function of putative
high-redshift clusters with photometric techniques alone.
\end{abstract}

\begin{keywords}
galaxies: clusters: general -- galaxies: high-redshift -- radio continuum:
general -- X-rays: galaxies -- X-rays: galaxies: clusters
\end{keywords}

\section{Introduction}

Clusters of galaxies are important probes for both cosmology and
astrophysics. They are the most massive virialised structures in the
Universe, and they are therefore excellent targets to study
large-scale structure formation. The cluster mass function directly
links to the normalisation of the power spectrum of density
perturbations, $\sigma_8$, which is the RMS in the density on a scale
of 8$h^{-1}$ Mpc ($h = H_0/100\,\rm km\,s^{-1}Mpc^{-1}$). The
evolution of the mass function with redshift reflects the growth
function of the Universe, which is determined by $\Omega_M$ (the
matter density parameter), $\Omega_{\Lambda}$ (the dark energy density
parameter), and $w$ (the dark energy equation-of-state
parameter). Further, clusters are of great interest to astrophysics
because they can be approximated by a `closed-box' environment. This
makes them ideal laboratories to study the interaction of galaxies and
the intergalactic medium, which is thought to play a large role in
galaxy formation and evolution. An important remaining question is why
so little baryonic matter in clusters forms stars (e.g. Cole 1991;
White \& Frenk 1991). Feedback processes coupling baryonic matter from
galaxies to the intergalactic medium are likely to counter further
condensation of the intergalactic gas and halt star-formation. Two
examples of these processes are galactic winds due to multiple
supernovae in star-burst galaxies (e.g. Heckman 2002) and outflow from
radio-loud AGN (e.g. Scannapieco \& Oh 2004; Fabian, Celotti \& Erlund
2006a).

Cluster searches have been carried out for decades. The first
extensive cluster catalogue was created by Abell (Abell 1958; Abell et
al. 1989). Unfortunately Abell only had single-filter photographic
plates at his disposal and therefore his work was severely complicated
by projection effects. Gladders et al. (1998) improved upon this
situation by showing that a redshift estimate can be determined by
using only two filters and targeting the bright, red elliptical galaxy
population in clusters which form the red sequence (e.g. Bower, Lucey
\& Ellis 1992). A particularly important advance in optical cluster
detection has come from the advent of large arrays of CCD detectors
allowing large surveys to be carried out efficiently, such as the
relatively shallow (z $<$ 0.4) but very wide-field Sloan Digital Sky
Survey (SDSS) (e.g Goto et al. 2002; Kim et al. 2002; Miller et
al. 2005). There have been numerous smaller-area surveys to higher
redshifts, for example the Palomar Distant Cluster Survey (Postman et
al. 1996), the ESO Imaging Survey (Lobo et al. 2000), and the Red
Sequence Cluster Survey (Gladders et al. 2005). Moving to even higher
redshifts, proto-clusters have been found by focussing on fields
around quasars or radio galaxies (e.g. Venemans et al. 2002) as a
large fraction of these objects have been shown to reside in clusters
(see Pascarelle et al. 1996; R\"ottgering et al.  1996). However,
selecting clusters in blank sky surveys with optical photometry at
redshifts $z \grtsim 1$ was impractical for a long time because of the
shifting of the 4000\,\AA\ break -- characteristic for the red,
passively evolving ellipticals predominantly found in clusters -- out
of the $I$-band to longer wavelength bands.

The launch of the {\it ROSAT} satellite in 1990 greatly advanced the
study of clusters in the X-ray regime, enabling the discovery of
hundreds of clusters up to $z \sim 1$. Later {\it Chandra} and {\it
XMM-Newton} provided the possibility to observe even deeper owing to
their unprecedented sensitivity and angular resolution. X-ray cluster
surveys now extend well above $z \sim 1$ (e.g. Stanford et al. 2001;
Rosati et al. 2004; Bremer et al. 2006). Stanford et al. (2006)
currently hold the record of the most-distant
spectroscopically-confirmed cluster at $z=1.451$. This cluster was
initially detected as an extended source in XMM X-ray data;
spectroscopic follow-up revealed five cluster galaxies within a
1$^{\prime}$-diameter. The temperature of the intra-cluster medium
(ICM) was shown to be $\sim 7$ keV, which is the highest detected in
any cluster at $z > 1$, implying a relatively massive cluster for such
a high redshift.

Another way to observe the ICM is through the Sunyaev-Zel'dovich effect
(S-Z effect, Sunyaev \& Zel'dovich 1970, 1972). This effect is visible
as a distortion of the Cosmic Microwave Background (CMB): as the
CMB-photons travel through the ICM they are subjected to inverse
Compton scattering, which shifts the energies of a small fraction of
the photons slightly upwards. In the last decade many detections have
been made of the S-Z effect (e.g. Birkinshaw, 1999; Carlstrom et al. 2000,
Jones et al. 2005) and comprehensive S-Z surveys are underway (Kneissl
et al. 2001). Cotter et al. (2002) used pointed S-Z observations to
detect hot gas associated with a $z \sim 1$ cluster of radio sources.

Recent development of wide-field infrared cameras, such as the
Wide-Field CAMera (WFCAM) on the United Kingdom InfraRed Telescope
(UKIRT), have brought photometric cluster surveys back into vogue by
enabling deep, large-scale infrared surveys and pushing the limit of
photometric cluster selection to significantly higher redshifts. The
UKIRT Infrared Deep Sky Survey (UKIDSS, Lawrence et al. 2006) is a
suite of both wide and deep infrared imaging surveys using WFCAM. The
Ultra Deep Survey is the deepest of these, and covers 0.8 deg$^2$ in
the Subaru XMM-Newton Deep Field (SXDF, Sekiguchi et al. [in prep.])
with a planned limiting magnitude of $K_{\rm Vega} = 23$. Deep
$BVRi'z'$ imaging data from the Subaru Telescope are also available on
the SXDF (Furusawa et al. [in prep.]), as is X-ray data by XMM-Newton
(Watson et al. 2004), radio data from the Very Large Array (Simpson et
al. 2006, Ivison et al. 2007), and Spitzer infrared data ranging from
3.6 to 24 micron (Lonsdale et al. 2005).

We previously exploited these exceptional multi-wavelength datasets to
execute a high-redshift cluster survey, using photometric redshifts,
as reported in Van Breukelen et al. 2006 (VB06). In VB06 we isolated
13 cluster candidates at $0.6 < z < 1.4$ in 0.5 deg$^2$ of the
SXDF. The highest-redshift cluster candidate, CVB13 (RA =
02$^h$18$^m$10.5$^s$, Dec = -05$^\circ$01$^\prime$05$^{\prime\prime}$
[J2000]), was estimated to lie at $z=1.39 \pm 0.07$.

In this paper we present the spectroscopic follow-up of the putative
CVB13 cluster with the DEep Imaging Multi-Object Spectrograph (DEIMOS,
Cowley et al. 1997) on the 10-metre Keck 2 telescope. Section 2
describes the observations and data reduction. The analysis and
discussion on the nature of the system are given in Section 3. Section
4 contains concluding remarks. Throughout this paper we use $H_0 =
70\, {\rm km\,s^{-1} Mpc^{-1}}$ ($h=0.7$), $\Omega_{\rm M} = 0.3$, and
$\Omega_{\Lambda} = 0.7$ and a $\Lambda$CDM power spectrum with
$\sigma_8=0.75$; all sky coordinates are quoted in equinox J2000.

\section{The data}

\subsection{Selection of the targets}
We selected four samples of target galaxies to create a Multi-Object
Spectroscopy (MOS) mask. Each of these samples comprised groups of
different priorities. The samples are described below; the masks were
created by allocating the maximum number of slits to sample 1,
priority 1, and moving down through all the priorities within the
sample before allocating slits to the next sample. 

{\bf Sample 1:} This sample included all galaxies that were determined
to be a possible cluster member of CVB13 by the algorithm described in
VB06. The algorithm presented in this paper used two methods to
detect clusters: Voronoi Tesselations and Friends-of-Friends. The
galaxies that were assigned to the cluster by both methods composed
the highest-priority target group within this sample. The second
priority group comprised the cluster members as selected only by
Friends-of-Friends, followed by the galaxies found only by Voronoi
Tesselations, as this is a less reliable method to determine cluster
membership (see VB06). The target list of this sample consisted of a
total of 125 objects, of which 18 objects of priority 1, 3 of priority
2, and 104 of priority 3. Slits could be placed on 61 of these 125
targets, divided into priorities 1: 10 galaxies, 2: 0 galaxies, and 3:
51 galaxies.

{\bf Sample 2:} By rotating the DEIMOS instrument by 90 degrees we
were able to utilise the long axis of the field-of-view to target
another candidate cluster from VB06 within the same observations. This
was cluster 6 (CVB6) at $z=0.76 \,\pm\,0.12$, RA $= 02^h18^m32.7^s$, Dec
= $-05^{\circ}01^{\prime}04^{\prime\prime}$. The target galaxies for
this sample were selected and prioritised exactly as for sample
1. There were 152 targets in total, divided into priorities 1: 30, 2:
1, 3: 121. Of these, 64 sources could be observed: 13 of priority 1, 1
of priority 2, and 50 of priority 3.


{\bf Sample 3:} The third sample consisted of various active galaxies
is the field, detected by their X-ray activity, radio emission, or
excess flux in the Spitzer 24-$\mu \rm m$ band. We found 48 radio sources
in the field, which made up priority group 1. The second priority
group included 31 X-ray sources. The last priority group consisted of
12 objects with a detection in the 24-$\mu \rm m$ band and a colour of
$J-K = 2$. The target list comprised 91 sources, of which 26 were
observed (15 radio sources, 8 X-ray sources, 3 24-$\mu \rm m$ sources).

\begin{figure}
\hspace{-0.7cm}
\includegraphics[height=58mm]{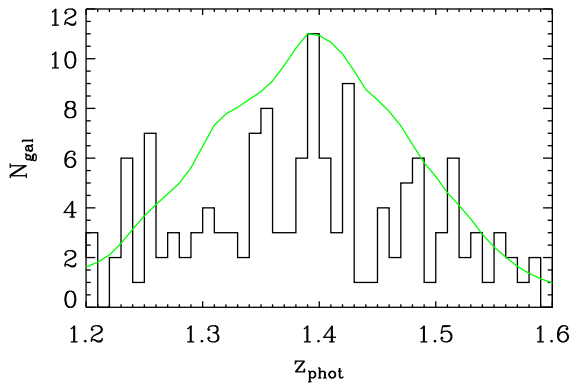}
\vspace{-0.2cm}
\caption{\small Histogram of the photometric redshifts of galaxies in
  the field of CVB13 ($02^h17^m54.6^s \leq {\rm RA} \leq
  02^h18^m26.4^s$ and $-05^{\circ}03^{\prime}05^{\prime\prime} \leq
  {\rm Dec} \leq -04^{\circ}59^{\prime}05^{\prime\prime}$ in the
  redshift range $1.2 < z < 1.6$. At $z=1.4$ the projected region of
  interest has proper size $3.6 \times 2.0 \rm\,Mpc^2$. The photometric
  redshifts were calculated using Bruzual \& Charlot (2003) modelled
  spectral energy distributions from stellar population synthesis (see
  VB06 for a discussion of the photometric redshifts). The solid line
  is the sum of the redshift probability functions (normalised to the
  peak of the histogram) of the galaxies with a photometric redshift
  within the shown redshift range.}
\label{photohist}
\end{figure}

{\bf Sample 4:} The final sample was designed to fill up all remaining
slits in the two masks. It included all galaxies in the field with a
photometric redshift of $1.25 < z_{phot} < 1.54$. Fig.~\ref{photohist}
is a histogram of the photometric redshifts of these objects;
overplotted is the sum of their redshift probability distribution
functions. The chosen redshift range corresponds to the $\pm$ 2-sigma
range on the photometric redshift of the cluster candidate CVB13. This
error on the cluster redshift reflects the combined redshift
probability distribution functions of the cluster candidate members as
given in VB06. This sample consisted of 1139 objects of which 43 were
observed.

\subsection{Observations and data reduction}

Optical spectroscopy was undertaken on the 21st of December 2006 using
the DEIMOS instrument on the Keck 2 telescope. The telescope was
pointed at RA = 02$^h$18$^m$20.00$^s$, Dec =
-05$^\circ$01$^\prime$04.9$^{\prime\prime}$ and the field-of-view of
the instrument, with a position angle of 90$^\circ$, was $16.7^\prime
\times 5^\prime$ in RA $\times$ Dec. The two MOS masks comprised
one-arcsecond wide slits. The 600 line mm$^{-1}$ grating was used in
conjunction with the GG495 order-blocking filter. A long-slit
observation of Feige 110 was taken for spectrophotometric
calibration. Four 1800-s integrations were taken for each mask, and
short-exposure images were taken through the mask after each
spectroscopic integration to ensure mask alignment. Seeing, as
measured on the alignment images, varied between 0.6$^{\prime\prime}$
and 1.0$^{\prime\prime}$. Conditions were photometric throughout.

The two-dimensional spectra were reduced with the DEEP2 DEIMOS Data
Pipeline\footnote{http://astro.berkeley.edu/\~{}cooper/deep/spec2d/}. The
pipeline processes the flats and arcs to determine the position of the
slitlets on the CCD and to find the two-dimensional wavelength
solution for each slitlet. Subsequently the spectra are flat-field
corrected and the curvature of the spectra in the spatial direction is
rectified. The separate science exposures are sky-subtracted and
finally combined into one mean, cosmic-ray rejected, science frame.

We extracted the one-dimensional spectra with a boxcar extraction
routine, using a 0.8$^{\prime\prime}$-width aperture. To enable flux
calibration, the standard star longslit exposure was reduced with the
IRAF package `twodspec'. The sensitivity function was determined using
the IRAF task `sensfunc' and the one-dimensional science spectra were
calibrated with the task `calibrate'. We note that for the average
seeing of 0.8$^{\prime\prime}$ we missed $\sim 13$ per cent of a
typical galaxy's flux within our aperture.

\subsection{The spectroscopic sample}
\begin{figure*}
\includegraphics[width=180mm]{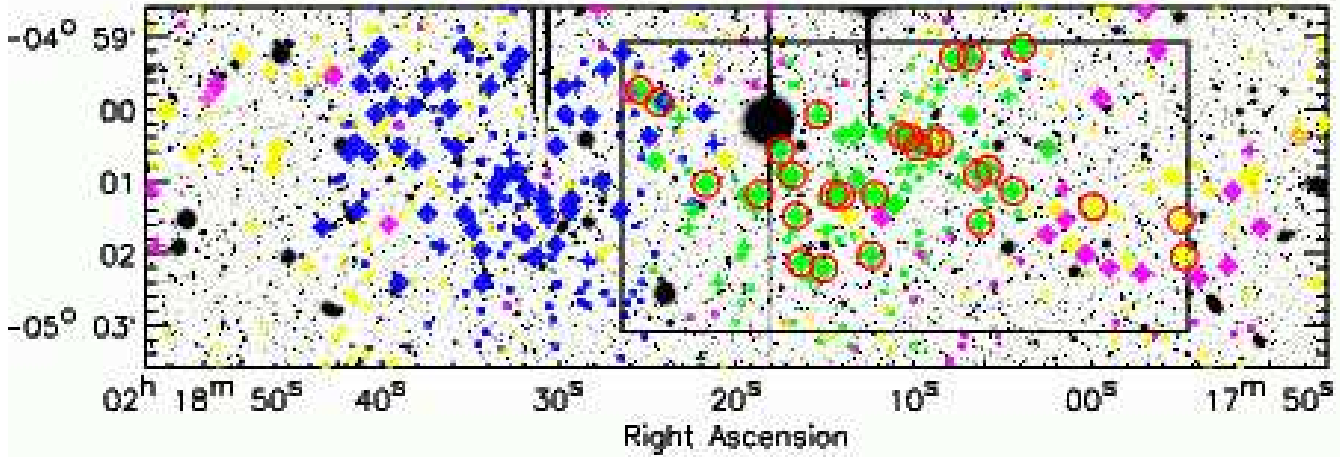}
\caption{\small Spectroscopy targets overlaid on the $i^{\prime}$-band
  image (Furusawa et al. [in prep.]) of the DEIMOS
  field-of-view. Sample 1 is depicted in green symbols, sample 2 in
  blue, sample 3 in purple, and sample 4 in yellow (see Section
  2.1). The small squares symbolise all the objects in the target
  list, the plusses are the objects that were actually observed, and
  the diamonds are all the targets for which a redshift could be
  determined. The red circles are the objects which are included in
  the final spectroscopic sample for this paper: they fall within the
  depicted box of $7^{\prime} \times 4^{\prime}$ centred on RA =
  02$^h$18$^m$10.5$^s$, Dec =
  -05$^\circ$01$^{\prime}$05$^{\prime\prime}$ and have redshifts of
  $1.25 < z < 1.54$ (see Section 2.3).}
\label{targets}
\end{figure*}

Overall the two MOS masks targeted 194 objects. Of these, we were able
to assign redshifts to 139 galaxies (72 per cent), which were divided
over the samples as follows: 38 redshifts in sample 1, 57 redshifts in
sample 2, 23 redshifts in sample 3, and 21 redshifts in sample 4. The
remaining 28 per cent of the observed galaxies had continua too faint
to allow a redshift determination from absorption features and showed
no emission lines. Fig.~\ref{targets} shows the positions of the
targets within the DEIMOS field-of-view where each sample is depicted
in a different colour.

In this paper we focus solely on cluster candidate CVB13. We therefore
imposed two selection criteria to create our final spectroscopic
sample: i) the objects had to be within the $7^{\prime} \times
4^{\prime}$ field centred on RA = 02$^h$18$^m$10.5$^s$ and Dec =
-05$^\circ$01$^{\prime}$05$^{\prime\prime}$ (see Fig.~\ref{photohist}
caption for RA and Dec limits of the sample); ii) the redshift range
was restricted to $1.25 < z < 1.54$, which was the original 2-sigma
range of cluster candidate redshift (see Section 2.1). There were 30
objects in our DEIMOS data that satisfied these criteria, of which 25
were from target-sample 1 and 5 from target-sample 4. These objects
are shown in Fig.~\ref{targets} with red circles. We added one galaxy
to our final sample that was observed with the Gemini Multi-Object
Spectrograph (GMOS, Hook et al. 2004) on the Gemini North Telescope in
Hawaii and satisfied the same selection criteria. The respective
sample selection and data description can be found in Van Breukelen et
al. (in prep.). This brought the total of the final spectroscopic
sample to 31 objects.

\section{Analysis and Discussion}

\subsection{Properties of the line emitters}

Due to the high redshift of our spectroscopic sample of 31 galaxies,
the continua of the spectra are weak, rendering redshift determination
by absorption features alone unfeasible. All redshifts in the final
sample were therefore calculated from the \oii~emission
line at 3727\,\AA. The identification of the \oii~line is
reliable because in some cases the doublet was resolved, in others
there were supporting absorption features, or there were no other
emission lines visible within the spectral range that would be
expected if the \oii~line was actually another emission
line at a different redshift. The one- and two-dimensional spectra of
the detected emission lines are shown in Fig. 3, where the
identified line features are labelled and noisy sky regions are
shaded. The three-dimensional distribution of the galaxies within our
target area is depicted in Fig.~\ref{3D} in comoving coordinates (X
corresponds to RA, Y to Dec and z to redshift). The galaxies are
colour-coded to redshift, from blue at $z=1.25$ to red at $z=1.53$.

\begin{figure*}
\begin{minipage}{175mm}
\begin{center}
\vspace{-0.2cm}	     
\epsfig{figure=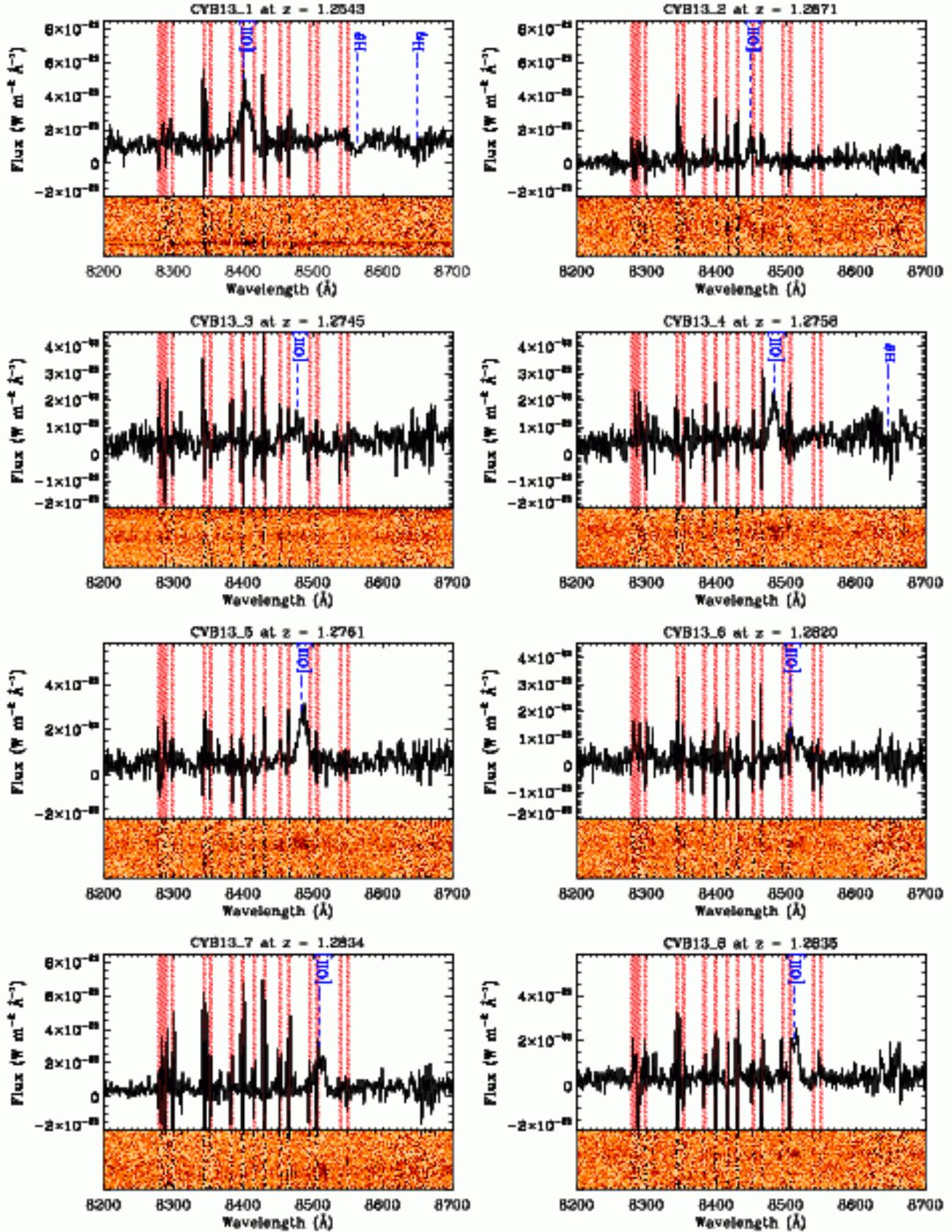,width=175mm}
\end{center}						     
\caption{\small One-dimensional and two-dimensional DEIMOS spectra of the
  galaxies in the field of CVB13. Identified emission and absorption
  features are labelled. The shaded regions denote noisy parts of the
  spectra due to sky lines.}
\end{minipage}						     
\end{figure*}						     
\begin{figure*}						     
\begin{minipage}{175mm}					     
\begin{center}						     
\vspace{-0.2cm}
\epsfig{figure=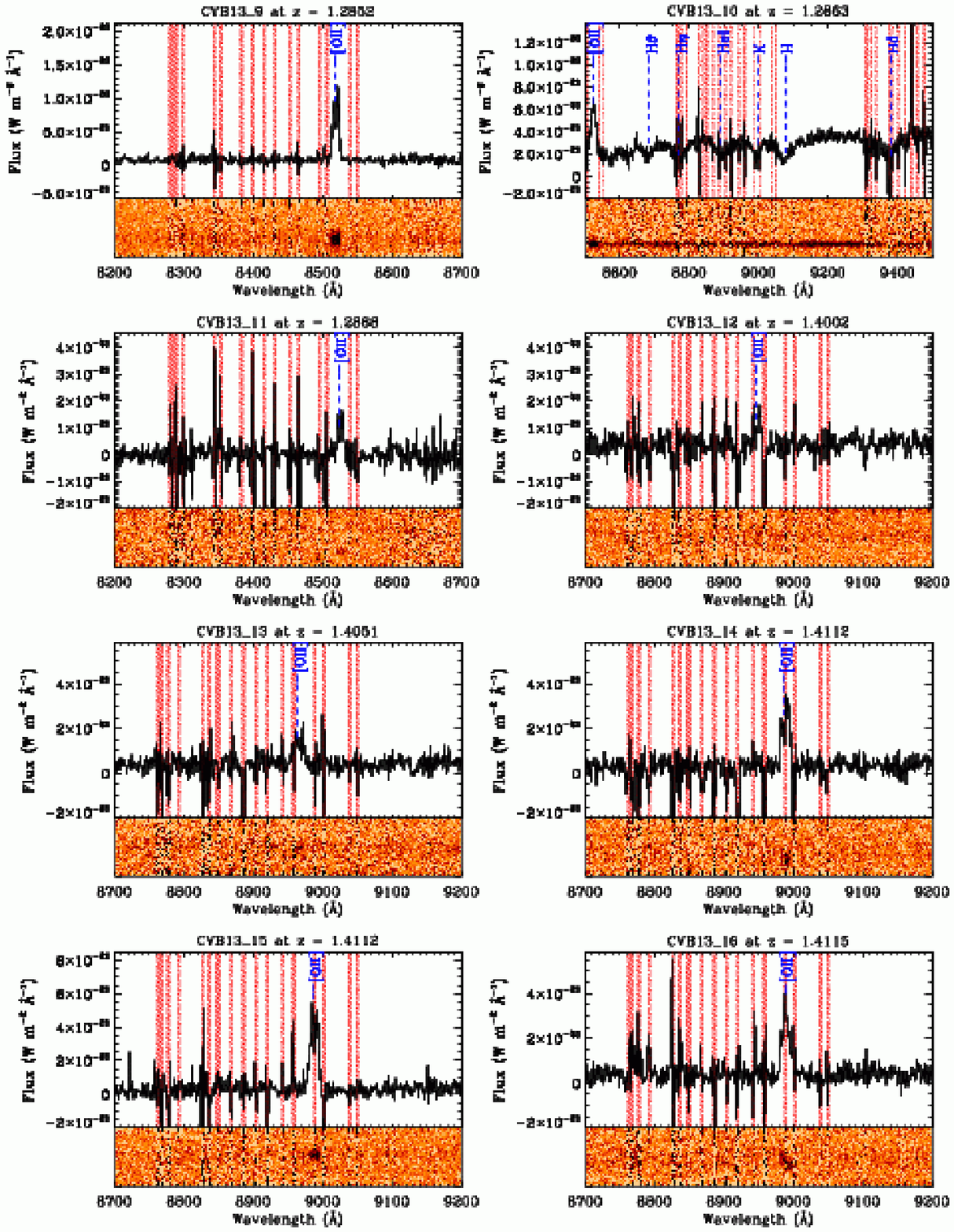,width=175mm}
\end{center}						      
\medskip
{\small {\bf Figure 3 cont'd.} One-dimensional and two-dimensional DEIMOS spectra of the
  galaxies in the field of CVB13. Identified emission and absorption
  features are labelled. The shaded regions denote noisy parts of the
  spectra due to sky lines.}
\end{minipage}
\end{figure*}
\begin{figure*}
\begin{minipage}{175mm}
\begin{center}
\vspace{-0.2cm}
\epsfig{figure=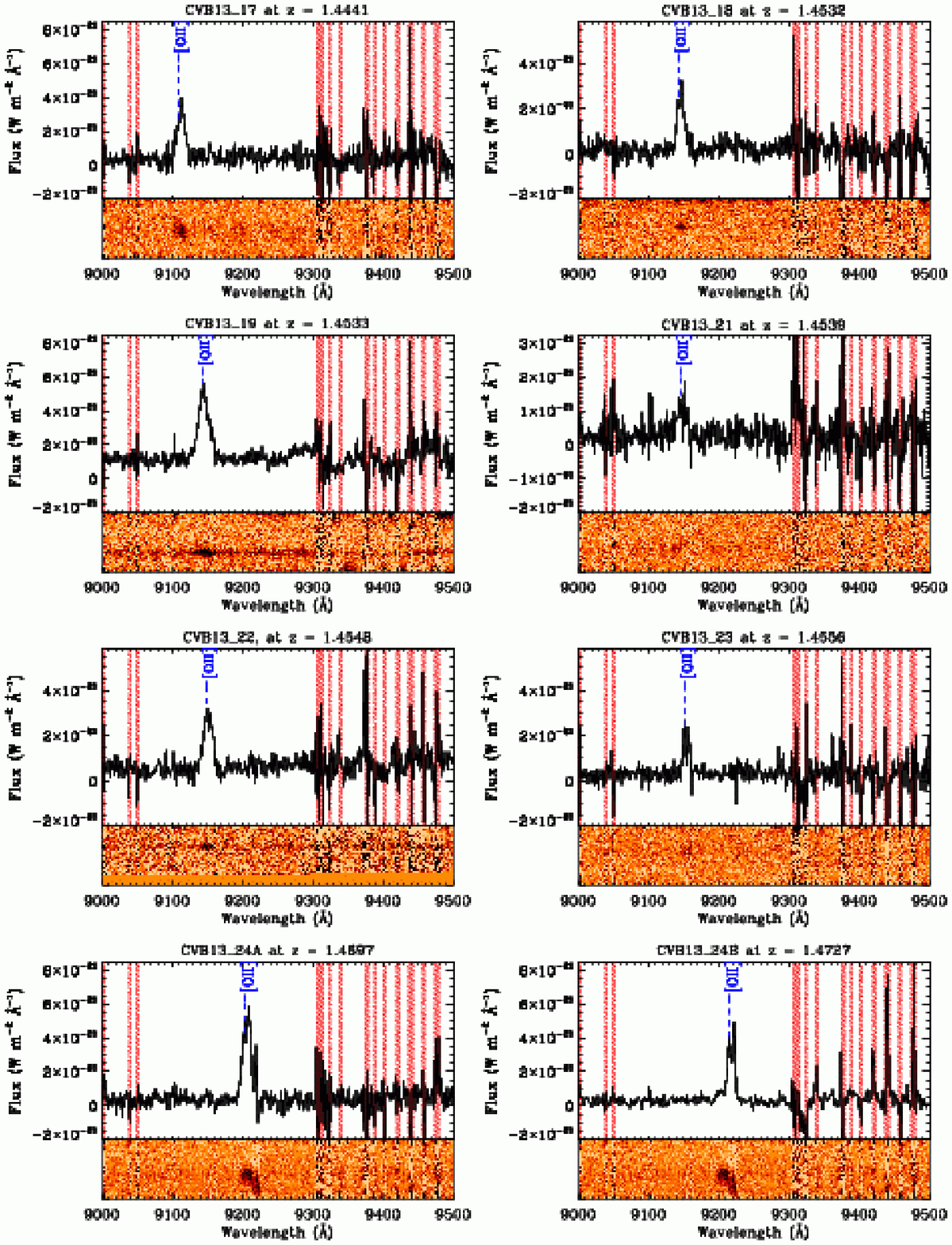,width=175mm}
\end{center}
\medskip
{\small {\bf Figure 3 cont'd.} One-dimensional and two-dimensional DEIMOS spectra of the
  galaxies in the field of CVB13. Identified emission and absorption
  features are labelled. The shaded regions denote noisy parts of the
  spectra due to sky lines.}
\end{minipage}
\end{figure*}
\begin{figure*}
\begin{minipage}{175mm}
\begin{center}
\vspace{-0.2cm}
\epsfig{figure=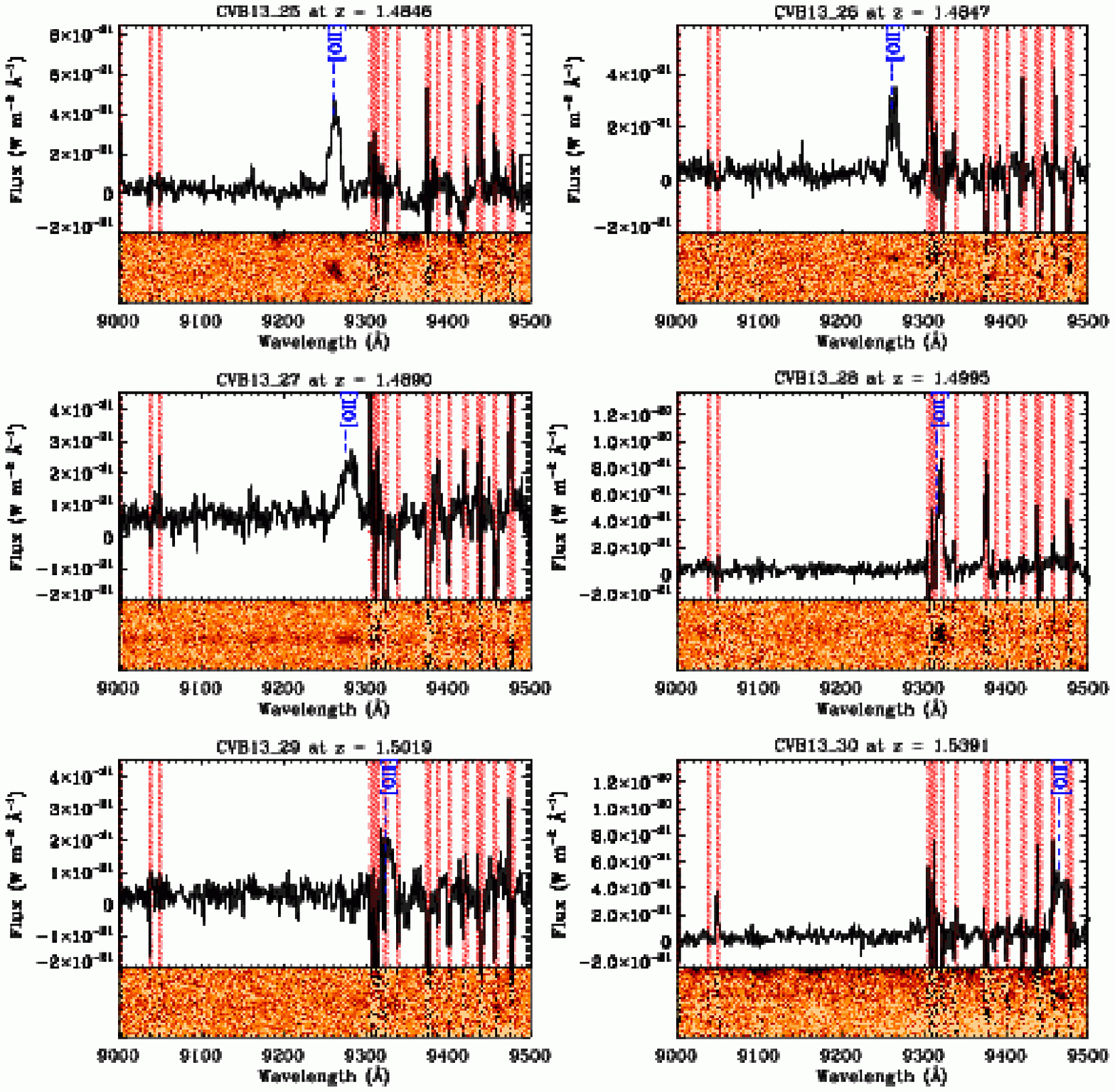,width=175mm}
\end{center}
{\small {\bf Figure 3 cont'd.} One-dimensional and two-dimensional DEIMOS spectra of the
  galaxies in the field of CVB13. Identified emission and absorption
  features are labelled. The shaded regions denote noisy parts of the
  spectra due to sky lines.}
\end{minipage}
\label{spec}
\end{figure*}

\begin{table*}
\begin{tabular}{lcccccrccr}
\hline
\noalign{\smallskip} 
\multicolumn{1}{c}{ID} & 
\multicolumn{1}{c}{RA} &  
\multicolumn{1}{c}{Dec.} &  
\multicolumn{1}{c}{$z_{\rm spec}$} & 
\multicolumn{1}{c}{$F_{\rm [OII]}$} &  
\multicolumn{1}{c}{$L_{\rm [OII]}$} &  
\multicolumn{1}{c}{$EW_0$} &
\multicolumn{1}{c}{SFR} &
\multicolumn{1}{c}{$i^{\prime}$} &
\multicolumn{1}{r}{$L_K$} \\
\multicolumn{1}{c}{} &
\multicolumn{1}{c}{[h m   s]} &   
\multicolumn{1}{c}{[$^{\circ}$  $^{\prime}$  $^{\prime\prime}$]} &   
\multicolumn{1}{c}{} &
\multicolumn{1}{c}{[$10^{-20}~\rm W~m^{-2}$]} &
\multicolumn{1}{c}{[$10^{34}~\rm W$]} &
\multicolumn{1}{c}{[$\rm \AA$]} &
\multicolumn{1}{c}{[$\rm M_{\odot}~yr^{-1}$]} &
\multicolumn{1}{c}{} &
\multicolumn{1}{r}{[$L^*_K$]} \\
\noalign{\smallskip} 
\hline 
\noalign{\smallskip} 
CVB13\_1   &  02:18:04.411 &  -05:01:08.25   &  1.25427 & 4.8  &    4.4  $\pm$  0.4 &   17 &  6.2 $\pm$ 1.8  & 22.480 &   3.5 \\ 
CVB13\_2   &  02:18:07.815 &  -04:59:17.22   &  1.26712 & 1.0  &    1.0  $\pm$  0.4 &   25 &  1.3 $\pm$ 0.7  & 23.777 &   1.1 \\ 
CVB13\_3   &  02:18:06.449 &  -05:00:54.87   &  1.27453 & 1.2  &    1.1  $\pm$  0.4 &   11 &  1.5 $\pm$ 0.7  & 23.292 &   2.1 \\ 
CVB13\_4   &  02:18:05.804 &  -05:00:50.10   &  1.27580 & 2.3  &    2.2  $\pm$  0.4 &   16 &  3.1 $\pm$ 1.0  & 22.934 &   1.7 \\ 
CVB13\_5   &  02:18:06.826 &  -04:59:17.36   &  1.27605 & 2.8  &    2.7  $\pm$  0.4 &   17 &  3.8 $\pm$ 1.2  & 23.772 &   1.8 \\ 
CVB13\_6   &  02:18:09.768 &  -05:00:31.80   &  1.28199 & 1.1  &    1.0  $\pm$  0.1 &   11 &  1.5 $\pm$ 0.4  & 24.226 &   0.2 \\ 
CVB13\_7   &  02:18:15.069 &  -05:02:12.49   &  1.28338 & 2.5  &    2.4  $\pm$  0.1 &   19 &  3.3 $\pm$ 1.0  & 22.693 &   1.3 \\ 
CVB13\_8   &  02:18:14.116 &  -05:01:12.02   &  1.28350 & 2.2  &    2.1  $\pm$  0.1 &   23 &  3.0 $\pm$ 0.9  & 24.588 &   0.3 \\ 
CVB13\_9   &  02:18:08.568 &  -05:00:27.32   &  1.28517 & 10.1 &    9.9  $\pm$  0.1 &   51 & 13.8 $\pm$ 4.0  & 23.610 &   0.3 \\ 
CVB13\_10  &  02:18:17.524 &  -05:00:34.91   &  1.28629 & 9.8  &    9.6  $\pm$  0.1 &   20 & 13.4 $\pm$ 3.8  & 22.642 &   2.7 \\ 
CVB13\_11  &  02:18:16.661 &  -05:01:28.73   &  1.28676 & 1.5  &    1.5  $\pm$  0.1 &   48 &  2.1 $\pm$ 0.6  & 26.278 &   0.5 \\ 
CVB13\_12  &  02:18:16.402 &  -05:02:08.13   &  1.40024 & 1.3  &    1.5  $\pm$  0.5 &   18 &  2.1 $\pm$ 0.9  & 23.027 &   1.5 \\ 
CVB13\_13  &  02:18:25.450 &  -04:59:44.17   &  1.40508 & 1.5  &    1.8  $\pm$  0.5 &   11 &  2.6 $\pm$ 1.0  & 23.371 &   3.1 \\ 
CVB13\_14  &  02:18:24.377 &  -04:59:54.12   &  1.41119 & 3.5  &    4.2  $\pm$  0.1 &   46 &  5.9 $\pm$ 1.7  & 22.753 &   0.7 \\ 
CVB13\_15  &  02:18:15.406 &  -05:00:04.62   &  1.41120 & 5.6  &    6.9  $\pm$  0.1 &   69 &  9.6 $\pm$ 2.8  & 24.243 &   0.2 \\ 
CVB13\_16  &  02:18:14.512 &  -05:01:10.79   &  1.41148 & 4.0  &    5.0  $\pm$  0.1 &   38 &  6.9 $\pm$ 2.0  & 23.489 &   0.9 \\ 
CVB13\_17  &  02:17:59.932 &  -05:01:20.99   &  1.44405 & 4.3  &    5.6  $\pm$  0.1 &   39 &  7.9 $\pm$ 2.3  & 23.875 &   0.8 \\ 
CVB13\_18  &  02:17:54.806 &  -05:02:03.29   &  1.45319 & 3.6  &    4.7  $\pm$  0.1 &   87 &  6.6 $\pm$ 1.9  & 24.173 &   0.6 \\ 
CVB13\_19  &  02:18:12.424 &  -05:02:01.65   &  1.45331 & 7.4  &    9.7  $\pm$  0.1 &   23 & 13.6 $\pm$ 3.9  & 23.012 &   3.4 \\ 
CVB13\_20* &  02:17:57.251 &  -05:02:23.82   &  1.45349 & 1.6  &    2.1  $\pm$  0.1 &   62 &  2.9 $\pm$ 0.8  & 23.484 &   3.9 \\    
CVB13\_21  &  02:18:09.084 &  -05:00:27.58   &  1.45393 & 1.1  &    1.5  $\pm$  0.1 &   11 &  2.1 $\pm$ 0.6  & 23.398 &   0.5 \\ 
CVB13\_22  &  02:18:03.930 &  -04:59:08.76   &  1.45479 & 4.2  &    5.5  $\pm$  0.1 &   29 &  7.7 $\pm$ 2.2  & 23.600 &   1.0 \\ 
CVB13\_23  &  02:18:10.251 &  -05:00:24.12   &  1.45558 & 1.9  &    2.5  $\pm$  0.1 &   23 &  3.4 $\pm$ 1.0  & 24.448 &   0.4 \\ 
CVB13\_24A &  02:18:12.252 &  -05:01:11.44   &  1.46968 & 8.2  &    1.1  $\pm$  0.1 &   91 & 15.5 $\pm$ 4.4  & 23.134 &   0.8 \\ 
CVB13\_24B &  02:18:12.172 &  -05:01:11.65   &  1.47274 & 5.0  &    6.8  $\pm$  0.1 &   43 &  9.5 $\pm$ 2.7  & 23.134 &$\lesssim 0.3$ \\
CVB13\_25  &  02:18:06.292 &  -05:01:34.62   &  1.48464 & 5.7  &    7.9  $\pm$  0.1 &   85 & 11.1 $\pm$ 3.2  & 23.606 &   1.1 \\ 
CVB13\_26  &  02:18:18.700 &  -05:01:10.74   &  1.48472 & 3.2  &    4.4  $\pm$  0.1 &   43 &  6.2 $\pm$ 1.8  & 24.398 &   0.4 \\ 
CVB13\_27  &  02:18:16.859 &  -05:00:55.27   &  1.48896 & 4.0  &    5.6  $\pm$  0.1 &   24 &  7.9 $\pm$ 2.3  & 23.809 &   2.3 \\ 
CVB13\_28  &  02:18:10.725 &  -05:00:22.11   &  1.49951 & 7.4  &    1.1  $\pm$  0.1 &   57 & 14.6 $\pm$ 4.2  & 23.910 &   0.4 \\ 
CVB13\_29  &  02:18:21.719 &  -05:01:02.32   &  1.50186 & 3.3  &    4.7  $\pm$  0.6 &   58 &  6.5 $\pm$ 2.0  & 24.033 &   1.4 \\ 
CVB13\_30  &  02:17:54.986 &  -05:01:32.49   &  1.53905 & 5.8  &    8.9  $\pm$  0.6 &   29 & 12.4 $\pm$ 3.6  & 23.591 &   0.4 \\ 
\noalign{\smallskip} 				    
\hline
\end{tabular}

\caption{\small Properties of the \oii~emitters. Column 1 states the
  galaxy ID; 24B was discovered serendipitously in the slit of 24A and
  does not meet the $K$-limit of the parent sample. The RA and Dec
  given in columns 2 and 3 are in equinox J2000. The (heliocentric)
  redshift and line flux of the emitters (columns 4 and 5) are derived
  from a double Gaussian fit to the \oii~3727\,\AA\ line profile. The
  line luminosity is shown in column 6, and column 7 is the rest-frame
  equivalent width. Column 8 shows the star-formation rate derived
  from $L_{\rm \oii~}$ (following Kennicutt 1998) and column 9 is the
  $i^{\prime}$-band magnitude (Vega). Column 10 shows the $K$-band
  luminosity in units of $L^*$ at the galaxy's redshift, calculated
  with $M^*_K = -24.34$ (Lin et al. 2004) at $z=0$ and assuming
  passive evolution (following prescriptions of Bruzual \& Charlot
  [2003]) with a formation redshift of $z_{\rm form} = 4$.  The error
  on the flux measurement is $\sim 1 \times 10^{-21}\rm W\,m^{-2}$,
  although in regions dominated by sky lines ($1.25 \lesssim z
  \lesssim 1.28$, $1.37 \lesssim z \lesssim 1.41$, and $z \grtsim
  1.5$) the error is on average four times higher. The error on the
  redshift measurement is estimated at $5 \times 10^{-5}$. The error
  on the luminosity is the propagated flux error. The error quoted for
  the SFR is the error derived from Eq.~\ref{sfr} and the error on the
  luminosity added in quadrature. *Observed with GMOS (see Van
  Breukelen et al. [in prep.]).}
\label{table}
\end{table*}

\begin{figure*}
\includegraphics[height=60mm]{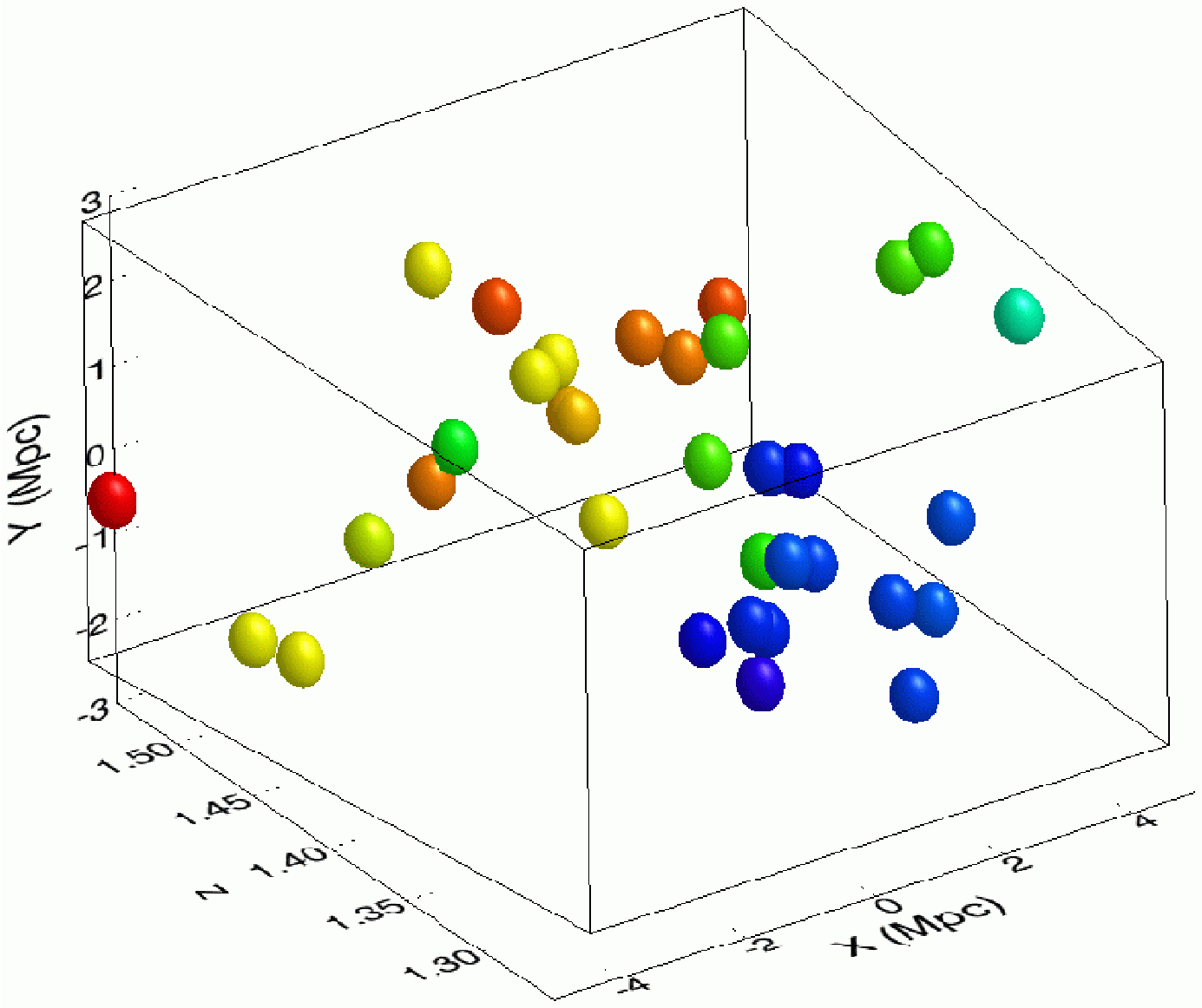}
\hspace{1cm}
\includegraphics[height=55mm,width=75mm]{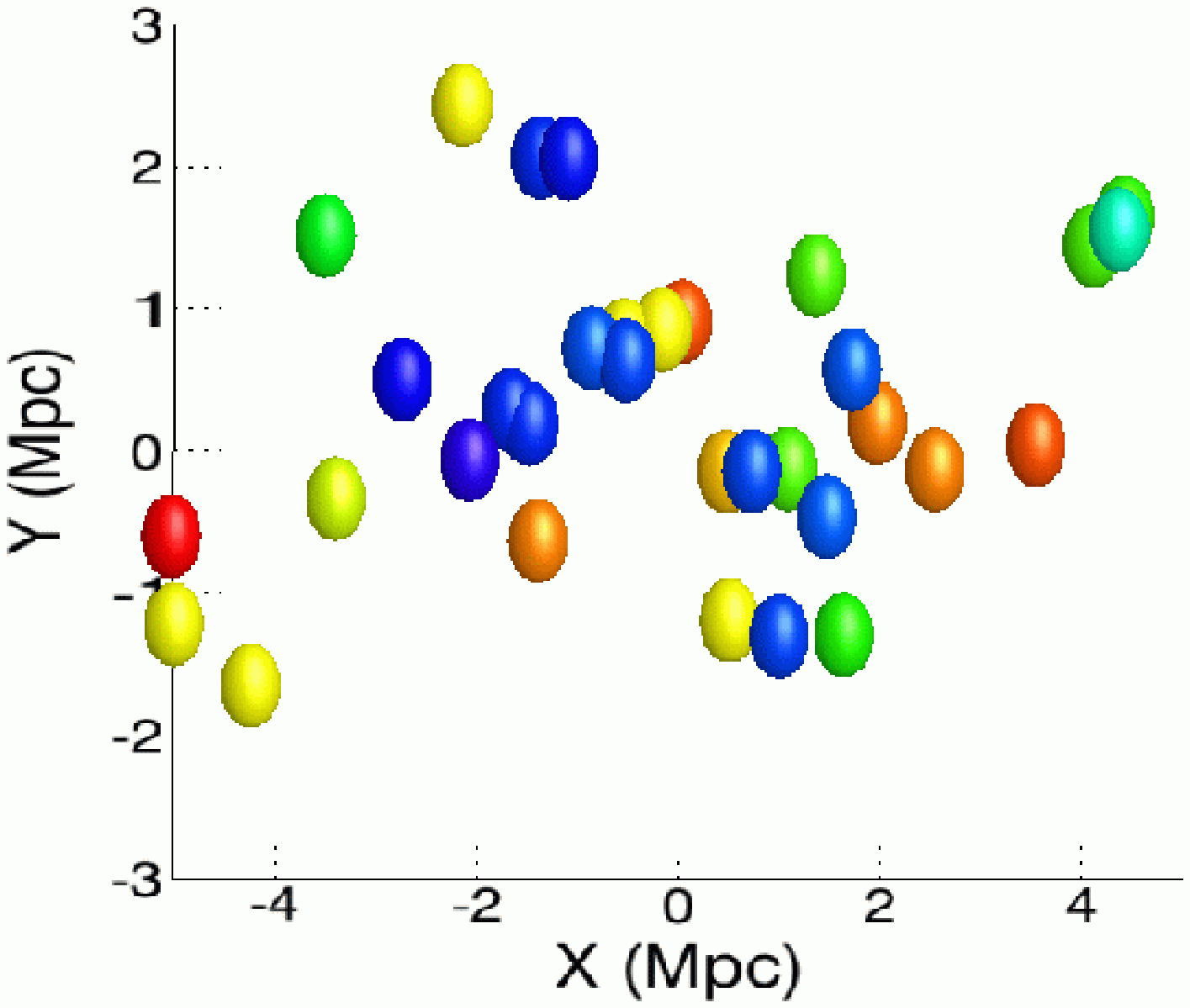}
\caption{\small The three-dimensional distribution of galaxies within
  our spectroscopic sample (see Fig.~\ref{targets}: red circles) in
  comoving coordinates (X corresponds to RA, Y to Dec and z to
  redshift). The plotted box reflects the field of CVB13 as defined in
  section 2.3; the front and back faces are defined by the RA and Dec
  limits as described in the caption of Fig.~\ref{photohist}. The
  galaxies are colour-coded to redshift, from blue at $z=1.25$ to red
  at $z=1.53$. {\it Right:} XY-projection of the galaxy distribution.}
\label{3D}
\end{figure*}

To determine the exact redshift, the line flux and the equivalent
width of the \oii~emitters, we fit a line profile to the
\oii~emission line in each object. As the
\oii$_{3727}$ line is a doublet, the fitted line profile
consists of two Gaussians at $\lambda_{rest} = 3726.1$\,\AA\ and
$\lambda_{rest} = 3728.8$\,\AA. The Full Width Half Maximum (FWHM) of
each Gaussian is assumed to be equal, and is a free parameter of the
fitted function. The other parameters are the redshift, the continuum
level, and the ratio of fluxes of the two lines. The flux ratio is
constrained to be within the limits for high and low density
(Osterbrock 1967), corresponding to $0.35 \leq F_{3729}/F_{3726} \leq
1.5$, where $F$ is the single line flux. Random errors on the flux and
redshift measurement are determined by fitting modelled lines with the
same procedure.

We calculate the Star-Formation Rate (SFR) of the galaxies using the
equation given by Kennicutt (1998):
\be
{\rm SFR} = (1.4 \pm 0.4) \times 10^{-34} L_{\rm \oii},
\label{sfr}
\ee where the SFR is in $\rm M_{\odot}\,yr^{-1}$ and $L_{\rm \oii}$ is
the luminosity of the \oii$_{3727}$ emission line in Watts. This
equation assumes $L_{\rm \oii}/L_{\rm H\alpha} = 0.57$ and a Salpeter
initial mass function with stellar masses between 0.1 and 100 $\rm
M_{\odot}$. Extinction or total obscuration is not corrected for and
therefore these estimates represent lower limits on the SFR. The
properties of the complete sample can be found in
Table~\ref{table}. In this table the redshift has been converted to
heliocentric redshift (see Danese, De Zotti \& Di Tullio 1980).

\subsection{Sample completeness}

To estimate the completeness of our spectroscopic sample, we have to
take three factors into account:

\begin{enumerate}

\item The original target list was compiled from the UKIDSS-UDS Early
Data Release (EDR) catalogue, which has a limiting $K$-band magnitude
of $K_{\rm Vega} = 20.6$. Cirasuolo et al. (2006) present a $K$-band
luminosity function of galaxies in the UKIDSS-UDS in different
redshift bins. We can calculate the total expected number of galaxies
using their Schechter function for redshift range $1.25 < z < 1.50$,
characterised by $\phi^* = 2.6 \,\pm\, 0.3 \rm \times 10^{-3}
Mpc^{-3}$, $M^*_{K,\rm Vega} = -24.93 \,\pm\, 0.16$, and $\alpha = -0.92\,
\pm\, 0.18$. In our field of $7^{\prime} \times 4^{\prime}$ and $1.25
< z < 1.54$ we would expect 391 galaxies in total; when we impose the
magnitude limit of $K_{\rm Vega} = 20.6$ we would only expect to
detect 99 objects, or 25 per cent. 

\item The nature of a MOS mask makes complete sampling of the target
  list impossible. Out of our original target list 177 objects
  satisfied our selection criteria (see Section 2.1), using
  photometric redshifts for the redshift criterion of $1.25 < z <
  1.54$. Of these objects, 75 were observed with the MOS masks. This
  means our spectroscopic data sampled 42 per cent of the galaxies we
  intended to target.

\item All the objects in our sample are \oii~emitters. We only detect
  the \oii~lines with a line flux $> 1.0 \times
  10^{34}$\,W\,m$^{-2}$. Furthermore, sky lines raise the flux limit
  of the \oii~line that we can detect in specific parts of the
  spectra. However, this affects lines with different line widths in
  different ways. We determine our detection completeness of
  \oii~lines with a flux greater than the flux limit of $1.0 \times
  10^{34}$\,W\,m$^{-2}$ by modelling a two-dimensional \oii$_{3727}$
  line. The modelled lines have a rest-frame FWHM varying between 1.3
  and 7.6\,\AA\ and fluxes from $10^{-20}$ to $10^{-19}$\,W\,m$^{-2}$,
  corresponding to the range of observed rest-frame FWHMs and line
  fluxes. We add the modelled lines at redshifts of $1.25 \le z \le
  1.54$ to a two-dimensional sky frame and try to recover the lines
  with the same procedure as used for the real spectra. Assuming a
  uniform distribution over the FWHM range and a Schechter
  distribution in flux, we calculate the fraction of missed sources at
  each redshift bin to find the detection completeness of \oii~lines
  with flux $> 1.0 \times 10^{34}$\,W\,m$^{-2}$. The result ranges
  from 22 per cent at the locations of the strongest sky lines up to
  97 per cent at the regions of the spectra with the least sky
  background noise.
\end{enumerate}

Fig.~\ref{prob} ({\it left}) shows the redshift distribution of the 31
objects in our sample. Overplotted on the spectroscopic-redshift
histogram is the expected number of \oii~emitters per redshift bin in
our field, calculated with the observed Schechter luminosity function
for \oii~ emitters at $z=1.47$ from Ly et al. (2006): $\log_{10}(\rm
\phi^* / Mpc^{-3}) = -1.97 \,\pm\, 0.06$; $\log_{10}(\rm L^* / W) =
35.60 \,\pm\, 0.05$; $\alpha = -0.78\, \pm\, 0.13$. The expected
number of \oii~emitters is corrected by multiplying by the
wavelength-dependent product of the three completeness factors, namely
$0.25 \times 0.42 ~\times$ (0.22 - 0.97). Note that this implicitly
assumes no correlation between \oii~flux and observed $K$-band
magnitude. However, if for example the rest-frame equivalent width of
the \oii~line is constant with the absolute blue magnitude of the
galaxies, we would expect a positive correlation between \oii~flux and
rest-frame blue magnitude. Therefore, if the colours of the
\oii~emitters are constant, we would expect a positive correlation
between \oii~flux and observed $K$-band magnitude. This would make the
completeness factor of 0.25 in point (i) an underestimate. On the
other hand, if the colour term is bluer for galaxies with a higher
star-formation rate -- and hence brighter \oii~emission -- then this
would counter the positive correlation. Studies of \oii~emission at $z
\sim 0 - 1$ (e.g. Hogg et al. 1998) do not allow us to separate these
effects, so we adopt the 0.25 completeness factor and discuss the
effects of extreme departures from this value in Section 3.3.

\subsection{Evidence of clustering}

\begin{figure*}
\vspace{-0.5cm}
\hspace{-0.5cm}
\includegraphics[height=60mm]{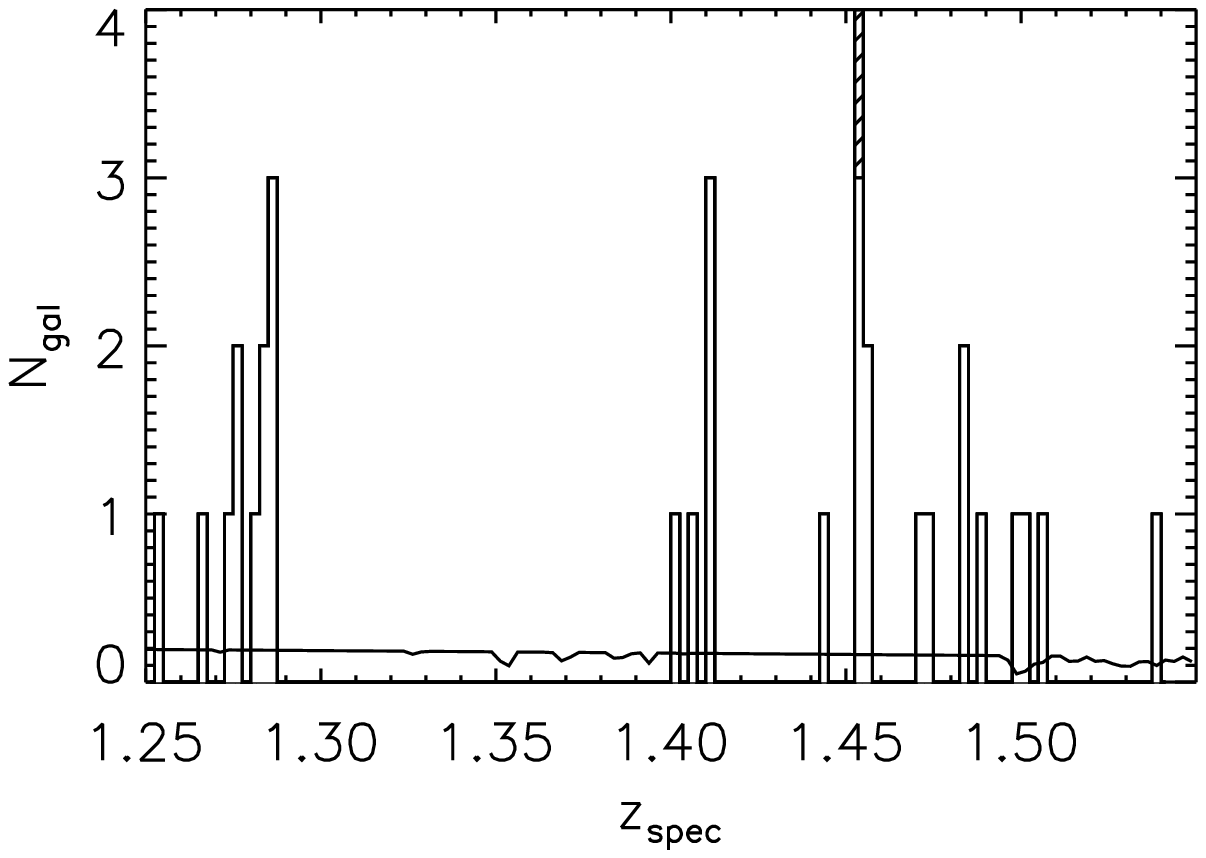}
\hspace{-0.5cm}
\includegraphics[height=60mm]{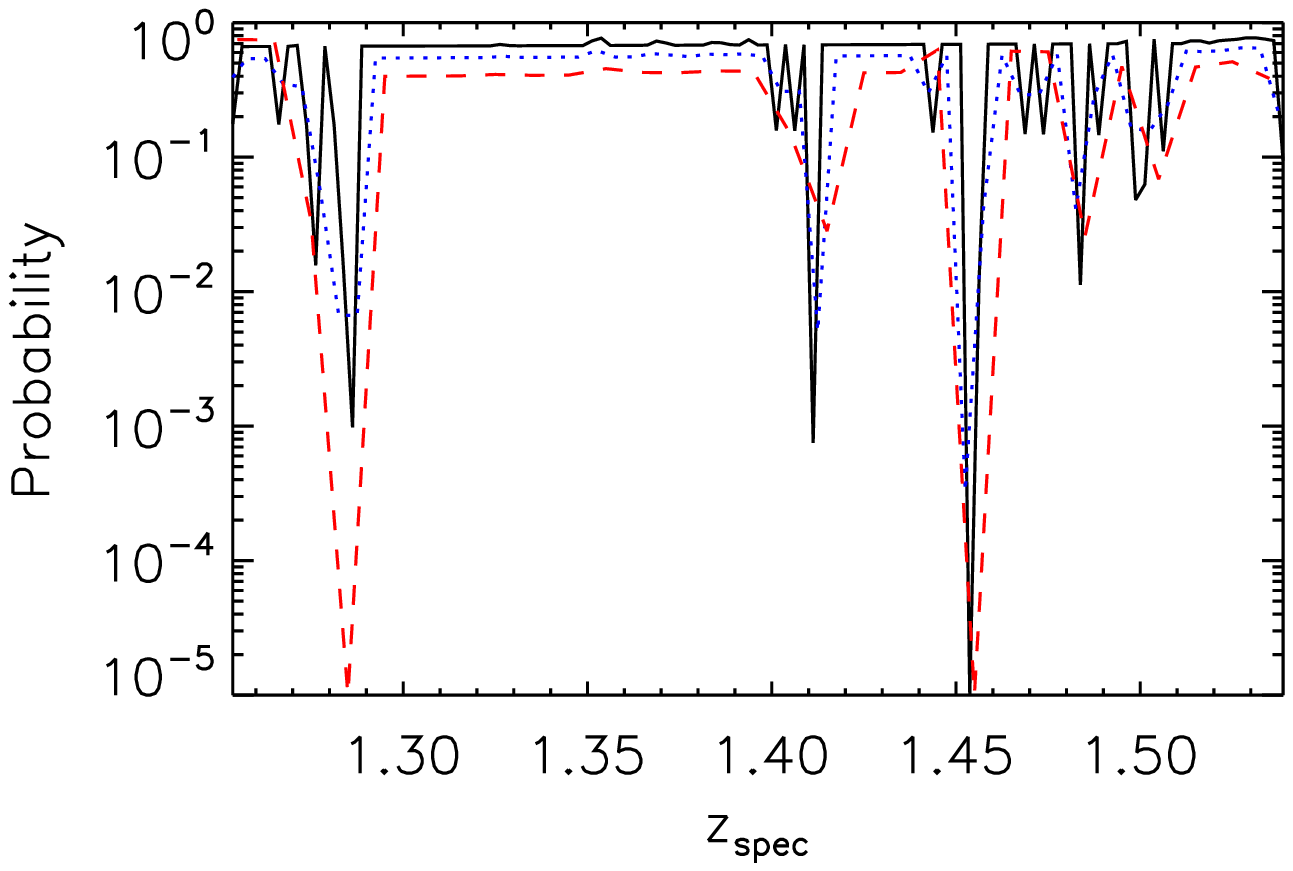}
\vspace{-0.2cm}
\caption{\small {\it Left:} histogram of the spectroscopic redshifts
  of the targeted galaxies with detected \oii~emission in the field of
  CVB13. The shaded bin represents the galaxy observed with
  GMOS. Overplotted is the expected number of \oii~emitters per
  redshift bin, taking into account our detection completeness (see
  section 3.2). The effects of the sky lines are somewhat smoothed in
  the plot owing to the binning of the histogram. {\it Right:} The
  probability of the detected number of \oii~emitters per redshift bin
  being due to the background model, using the method of Brand et
  al. (2003) described in Section 3.3. The solid black line is for a
  bin size of $\Delta z = 0.0025$, the dotted blue line for $\Delta z
  = 0.005$, and the dashed red line for $\Delta z = 0.01$.}
\label{prob}
\end{figure*}

The distribution of photometric redshifts in the field of CVB13 has a
clear peak at $z=1.4$, as is shown in
Fig.~\ref{photohist}. Spectroscopic follow-up however shows that the
single redshift-peak at $z=1.4$ separates into three less significant
peaks at $z = 1.40$, $z=1.45$, and $z=1.48$, as pictured in
Fig.~\ref{prob} ({\it left}). There is also a clear overdensity at
$z=1.28$ which was not detected in the photometric redshift catalogue.

Following the method of Brand et al. (2003) we calculate the
probability in each redshift bin (using Poissonian low-number
statistics) that the number of \oii~emitters is greater than the
number detected, given the number of \oii~emitters predicted by the
model. The result is shown in Fig.~\ref{prob} ({\it right}) for bin
sizes of $\Delta z = 0.0025$ (120 bins, black solid line), $\Delta z =
0.005$ (60 bins, blue dotted line), and $\Delta z = 0.01$ (30 bins,
red dashed line). For a bin size of $\Delta z = 0.01$ the redshift
peaks at $z=1.28$ and $z=1.45$ both have a probability of
$10^{-5}$. As there are 30 bins with this bin size, one would expect
to find such a low value in 0.03 per cent drawn from a set of random
realizations.  Based on these statistics we deem the structures at
$z=1.28$ and $z=1.45$ to be highly significant overdensities. Other
peaks in the redshift distribution could reflect genuine structures
but could alternatively be due to Poisson noise in the background
distribution. Note that if the completeness factor of point (i) in
Section 3.2 is smaller than 0.25, the overdensities would be even more
significant; if the factor is larger than 0.25 this would reduce the
probability of the two structures up to $\sim 0.01$ for a factor of
1. However, this large a completeness value is disproved by the
existence of object CVB13\_24B which has strong \oii~emission but $K >
20.6$, and by the occurence of similar high equivalent width objects
in the sample of Hogg et al. (1998).

\begin{figure*}
\includegraphics[height=55mm]{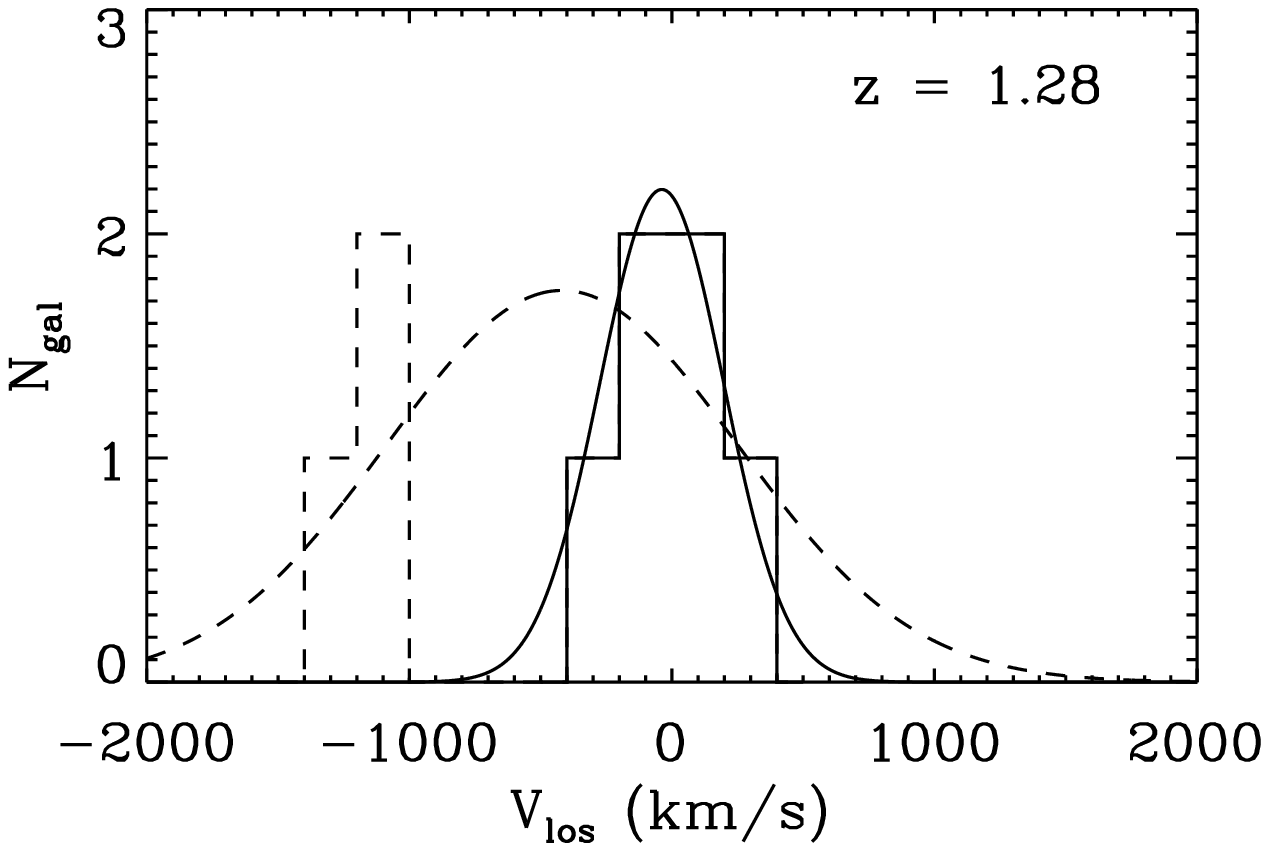}
\includegraphics[height=55mm]{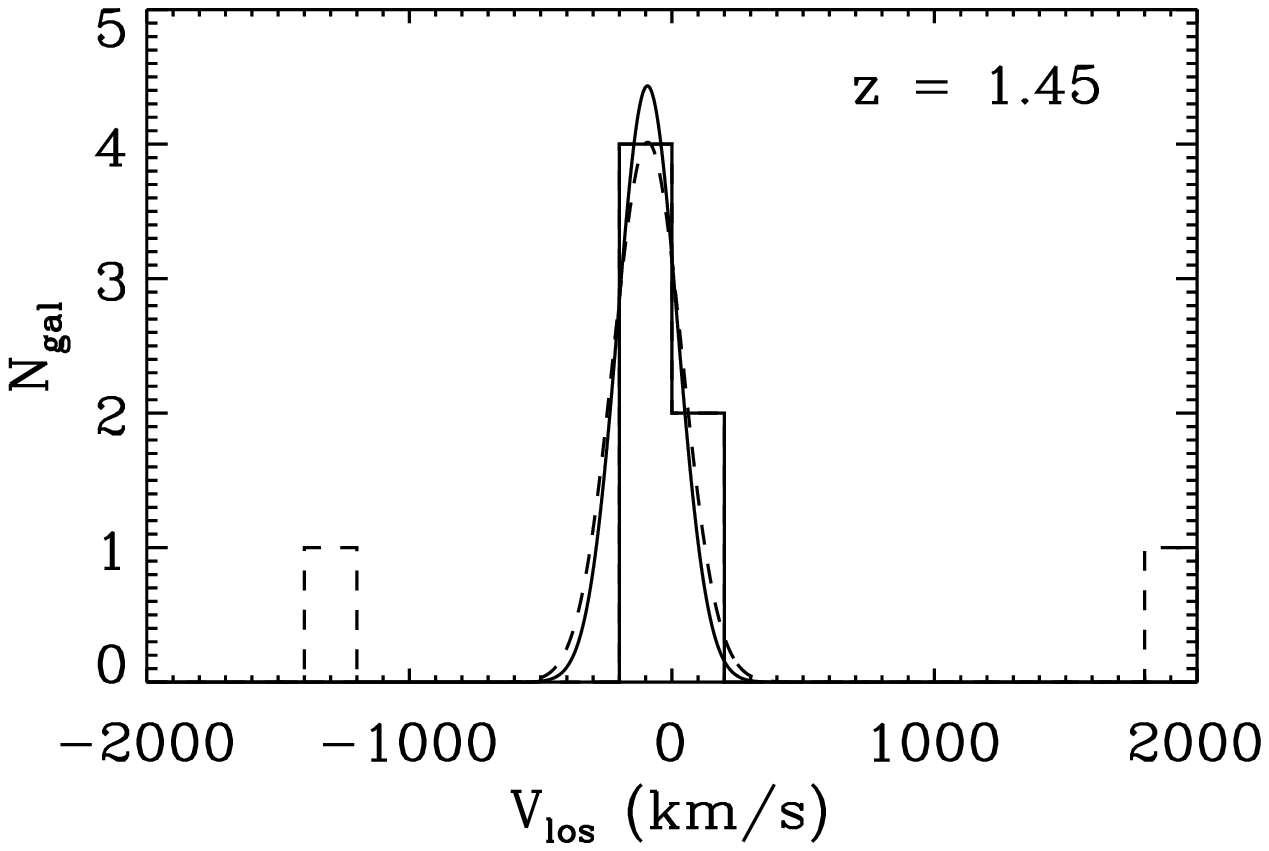}
\caption{\small {\it Left:} The line-of-sight velocity distribution
  for the structure at $z=1.28$. The solid lines in both the histogram
  and the fitted Gaussian comprise all galaxies within a range of
  $\pm$ 1000 km s$^{-1}$ from the centre. The dashed lines show the sample within
  $\pm$ 2000 km s$^{-1}$. {\it Right:} Velocity distribution for the structure at
  $z=1.45$. Line styles are as in the diagram on the left.}
\label{veldisp}
\end{figure*}

We do not detect any obvious signs of clustering in RA and Dec in our
spectroscopic sample (see Fig.~\ref{3D}). This is principally because
(i) we were unable to spectroscopically observe every photometric
candidate cluster galaxy within the allocated time (ii) the MOS
instrument was unable to completely sample the cluster core, as
slitlets cannot be placed within 1.5$^{\prime\prime}$ along the
spatial axis, nor next to each other along the dispersion axis as the
spectra will overlap, and (iii) we could only determine redshifts for
\oii~emitting galaxies and not for the red, passively-evolving
galaxies, which are customarily more clustered (e.g. Meneux et
al. 2006).

\subsection{Velocity dispersions and mass estimates}

The line-of-sight velocity distribution of the two statistically
robust structures at $z=1.28$ and $z=1.45$ is shown in
Fig.~\ref{veldisp}. The bi-weighted mean of the redshifts and the
`gapper' estimate of the velocity dispersion are calculated for each of
the structures, using the procedure outlined in Beers, Flynn \&
Gebhardt (1990). Subsequently, we estimate the masses of the groups
using the relation between cluster mass $M_{200}$ (the mass contained
within a sphere of radius $r_{200}$ for which the mean density is $200
\,\rho_{\rm cr}$) and virial velocity found by Evrard (2004)
(Eq.~\ref{mass}), substituting $\sigma_{\rm v} = \sqrt{3}\sigma_{v,{\rm
los}}$:

\be
M_{200} = \frac{10^{15}h^{-1}{\rm
    M_{\odot}}}{H/H_0}\Big(\frac{\sqrt{3}\sigma_{\rm v,los}}{1080\,{\rm km\,s^{-1}}}\Big)^3, 
\label{mass}
\ee
where for a flat Universe:
\be
H(z) = H_0\sqrt{\Omega_{\Lambda} + \Omega_{\rm M} (1+z)^3}.
\label{hz}
\ee

An alternative mass estimate is based on the stellar light of the
groups observed in the $K$-band. This is done in a number of
steps. First, we calculate the $K$-band luminosity of each galaxy in
terms of $L^*_K$ at the appropriate redshift. This parameter is
determined by assuming $M^*_K = -24.34$ for the cluster luminosity
function (Lin, Mohr \& Stanford 2004) at $z=0$, and passive evolution
of the $K$-band luminosity function given a formation redshift of
$z_{\rm form}=4$. Next, we sum all the luminosities of the observed
galaxies and correct for the missed fraction of light due to the flux
limit, according to the luminosity function of Lin et al. (2004). To
arrive at the total mass we then take a mass-to-light ratio of
$(M/{\rm M_{\odot}})/(L/{\rm L_{\odot}})$ of $75h$ (Rines et
al. 2001) which is assumed constant with redshift {\it in terms of
$L^*_K$} (note that assuming passive evolution of the luminosity
function implies $L^*_K$ itself increases with redshift). The mass
thus estimated is only a lower limit, as we do not correct for the
fact that our spectroscopic sampling of the group members is most
likely incomplete.

As is illustrated in Fig.~\ref{veldisp} ({\it left}) the mean redshift
and velocity dispersion of the structure at $z=1.28$ is highly
dependent on the limit imposed on the individual galaxy velocity for
group membership. If we assume all galaxies within a range of 2000
km\,s$^{-1}$ are members of the group, we obtain a mean redshift of
$\overline{z} = 1.282\,\pm\,0.002$ and a velocity dispersion of
$\sigma_{\rm v,los} = 668\,\pm\,149$ km s$^{-1}$. Placing the limit at
1000 km\,s$^{-1}$ however gives $\overline{z} = 1.2847\,\pm\,0.0008$,
$\sigma_{\rm v,los} = 236\,\pm\,63$ km\,s$^{-1}$. Assuming the group is
virialised, the dynamically inferred masses are $5 \times 10^{14}\rm
M_{\odot}$ and $3 \times 10^{13}\rm M_{\odot}$ respectively. The mass
estimates from the total stellar light are $8 \times 10^{13}\rm
M_{\odot}$ and $4 \times 10^{13}\rm M_{\odot}$ respectively for the
two galaxy samples.

The structure at $z=1.45$ gives similar velocity distributions for the
samples in the two velocity ranges (see Fig.~\ref{veldisp} [{\it
right}]). We arrive at $\overline{z} = 1.4542\,\pm\,0.0004$ and
$\sigma_{\rm v,los} = 113\,\pm\,57$ km\,s$^{-1}$; the inferred dynamical mass
is only $5 \times 10^{12}\rm M_{\odot}$. The mass derived from the
stellar light yields a much higher estimate of $9 \times 10^{13}\rm
M_{\odot}$. Since we have spectroscopic confirmation for six to eight
galaxies in the group with a total stellar mass of $1.6 - 1.8 \times
10^{12} \rm M_{\odot}$ we believe $\sim$ a few $10^{13} \rm M_{\odot}$
is a sensible minimum mass for the associated dark matter halo.

\subsection{X-ray and radio properties}
\begin{figure*}
\includegraphics[height=80mm]{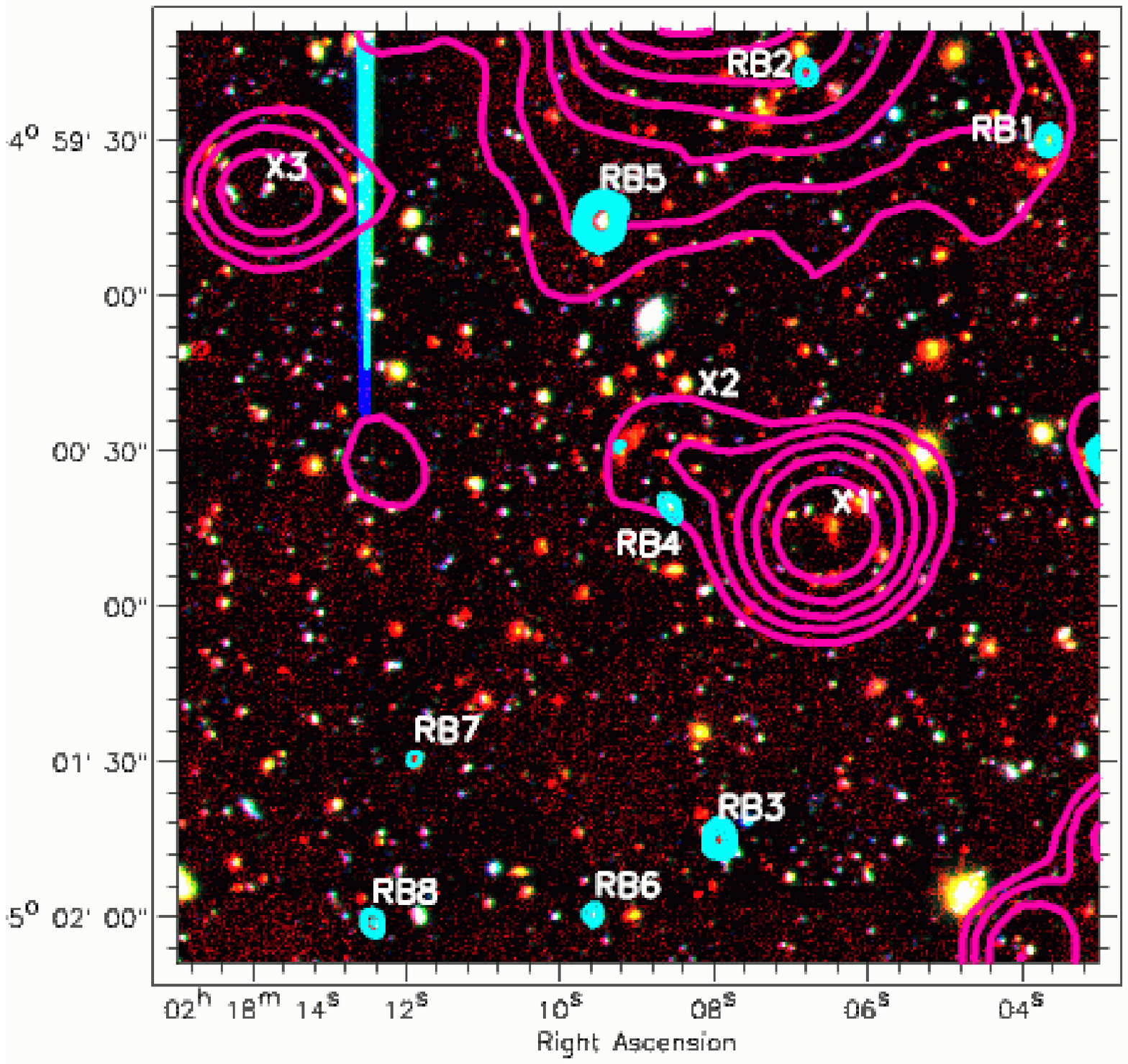}
\hspace{0.3cm}
\includegraphics[height=80mm]{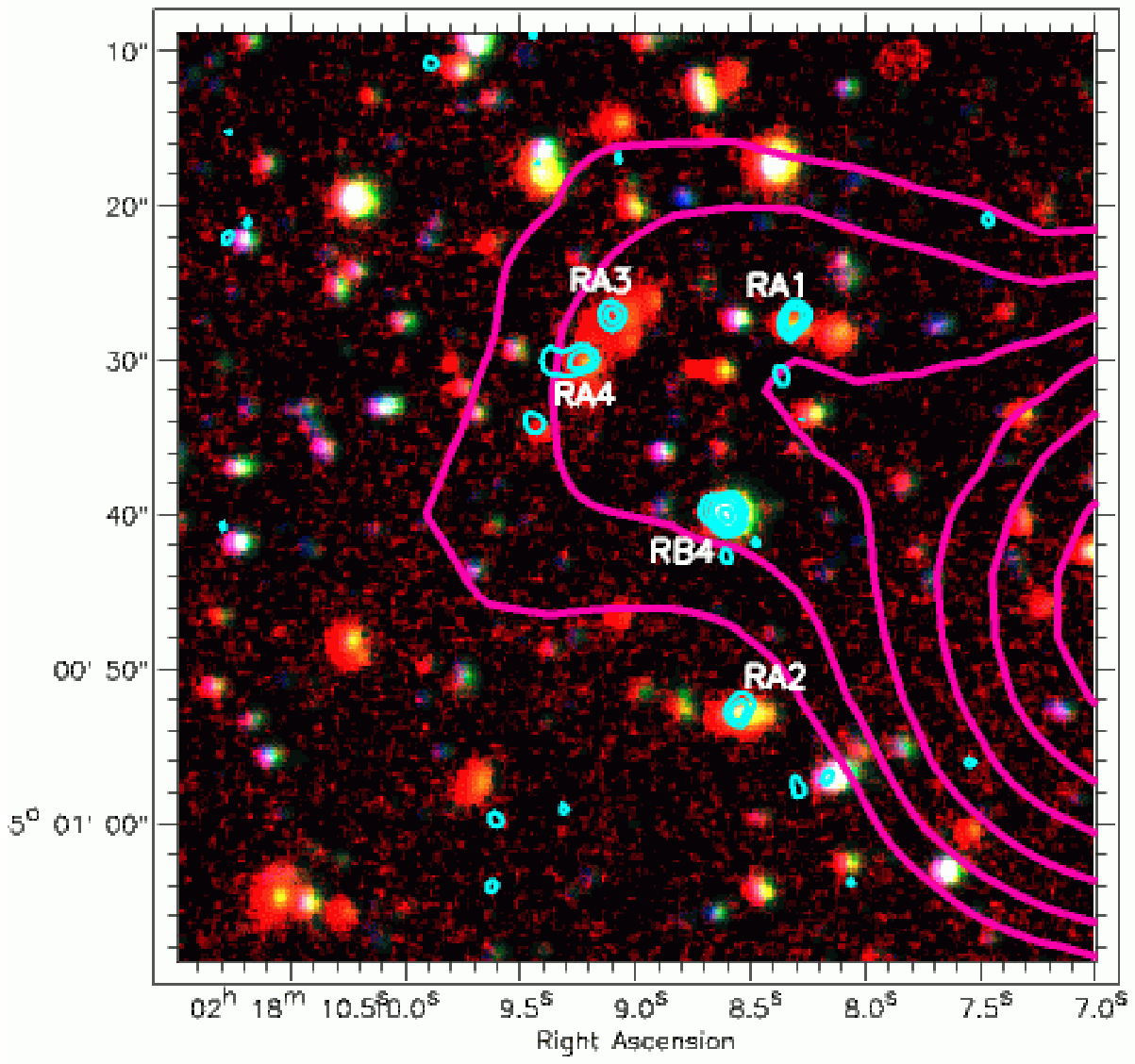}
\caption{\small {\it Left:} $Bi^\prime K$ colour image (Furusawa et
  al. [in prep.]) of the central $3^{\prime} \times 3^{\prime}$ region of
  CVB13. X-ray contours are from a signal-to-noise radio map in the
  0.5 - 2 keV band and are overlaid in purple (contour levels at
  [$\sqrt{2}\sigma, 2\sigma, 2\sqrt{2}\sigma, ...$]); the 1.4-GHz VLA
  B-array radio map is contoured in blue (contour levels at [$4\sigma,
  4\sqrt{2}\sigma,8\sigma, ...$]). The radio beam size is
  $5^{\prime\prime} \times 4^{\prime\prime}$ at PA =
  170$^\circ$. X-ray and radio sources are labelled as in
  Table~\ref{objects}. {\it Right:} Colour image as left, zoomed in on
  the central $1^{\prime} \times 1^{\prime}$ region. The radio
  contours are from the higher resolution VLA A-array radio map
  (contour levels at [$2\sqrt{2}\sigma,4\sigma, 4\sqrt{2}\sigma, ...$]),
  with a beam size of 1.9$^{\prime\prime}$ by 1.6$^{\prime\prime}$ at
  PA = 22$^\circ$.}
\label{overlay}
\end{figure*}

\begin{table*}
\begin{tabular}{lccrrrl}
\hline
\noalign{\smallskip} 
\multicolumn{1}{l}{ID} & 
\multicolumn{1}{c}{RA} &  
\multicolumn{1}{c}{Dec.} &  
\multicolumn{1}{r}{$F$} &  
\multicolumn{1}{r}{$z_{\rm photo}$} & 
\multicolumn{1}{r}{$z_{\rm spec}$} & 
\multicolumn{1}{l}{Ref.} \\
\multicolumn{1}{l}{} &
\multicolumn{1}{c}{[h m   s]} &   
\multicolumn{1}{c}{[$^{\circ}$  $^{\prime}$  $^{\prime\prime}$]} &   
\multicolumn{1}{r}{$\rm [W~m^{-2}]$} &
\multicolumn{1}{r}{} &
\multicolumn{1}{r}{} &
\multicolumn{1}{l}{} \\
\noalign{\smallskip} 
\hline 
\noalign{\smallskip} 
X1   &  02:18:06.52 &  -05:00:43.9  &   $2.5 \,\pm\, 0.3 \times 10^{-17}$ & 1.06  & -     &  \\
X2   &  02:18:08.37 &  -05:00:30.3  &	$9.1 \,\pm\, 2.1 \times 10^{-18}$ & -     & -     &  \\
X3   &  02:18:13.83 &  -04:59:39.2  &	$6.4 \,\pm\, 1.5 \times 10^{-18}$ & -     & 1.258 & Simpson et al. (in prep) \\
\hline
\noalign{\smallskip} 
\multicolumn{1}{l}{} &
\multicolumn{1}{c}{[h m   s]} &   
\multicolumn{1}{c}{[$^{\circ}$  $^{\prime}$  $^{\prime\prime}$]} &   
\multicolumn{1}{r}{[$\mu \rm Jy$]} &
\multicolumn{1}{r}{} &
\multicolumn{1}{r}{} &
\multicolumn{1}{l}{} \\
\noalign{\smallskip} 
\hline 
\noalign{\smallskip} 
RB1      &   02:18:03.66 &  -04:59:30.1  & $122  \,\pm\,26 $ & 1.05 & -      &              \\
RB2*     &   02:18:06.83 &  -04:59:16.8  & $174  \,\pm\,50 $ & 1.11 & 1.276  & This paper (CVB13\_5)   \\
RB3      &   02:18:07.94 &  -05:01:45.0  & $256  \,\pm\,27 $ & -    & -      &              \\
RB4      &   02:18:08.59 &  -05:00:40.0  & $102  \,\pm\,29 $ & -    & 0.493  & Simpson (priv. comm.)    \\
RB5      &   02:18:09.48 &  -04:59:45.5  & $2347 \,\pm\,28 $ & 1.35 & 1.094  & Simpson et al. (in prep) \\
RB6      &   02:18:09.55 &  -05:02:00.0  & $68   \,\pm\,22 $ & -    & 1.268  & Simpson (priv. comm.)    \\
RB7      &   02:18:11.89 &  -05:01:29.6  & $49   \,\pm\,21 $ & 0.75 & 0.918  & This paper   \\
RB8      &   02:18:12.44 &  -05:02:01.3  & $104  \,\pm\,27 $ & 1.32 & 1.453  & This paper (CVB13\_19)   \\
RA1      &   02:18:08.30 &  -05:00:27.2  & $67   \,\pm\,20 $ & 1.02 & -      &              \\
RA2      &   02:18:08.57 &  -05:00:52.8  & $73   \,\pm\,30 $ & 0.94 & -      &              \\
RA3      &   02:18:09.10 &  -05:00:27.2  & $66   \,\pm\,29 $ & 1.27 & 1.454  & This paper (CVB13\_21)   \\
RA4      &   02:18:09.23 &  -05:00:30.0  & $109  \,\pm\,38 $ & 1.32 & -      &              \\
\noalign{\smallskip} 				    
\hline
\end{tabular}

\caption{\small Table of the X-ray and radio sources in the
  $3^{\prime} \times 3^{\prime}$ central region of CVB13. The IDs
  correspond to the labels in Fig.~\ref{overlay}. The RA and Dec are
  given in columns 2 and 3. The values in column 4 are the X-ray flux
  (0.5 - 10 keV) for the X-ray sources, and the 1.4-GHz radio flux
  density for the radio sources. The photometric redshifts (column 5)
  were determined as described in VB06. The spectroscopic redshifts
  (column 6) are from various sources as referenced in column 7. *This
  object appears to be a resolved source in the B-array data, but the
  A-array data shows that it actually consists of 3 separate
  objects. The central object on its own has a flux of $F = 113\,\pm\,24\,\mu$Jy.}
\label{objects}
\end{table*}

We have performed an analysis of the X-ray emission around CVB13.
Fig.~\ref{overlay} ({\it left}) shows a three-colour ($Bi^{\prime}K$)
image of the central $3^{\prime} \times 3^{\prime}$ field with X-ray
and radio (VLA B-array) contours overlaid. We detect X-ray emission at
RA = $02^h18^m09^s$ and Dec =
$-05^{\circ}00^{\prime}30^{\prime\prime}$ (object X2, see
Table~\ref{objects} and Ueda et al. [in prep] source no. 0712) which
is coincident with the optical/near-infrared cluster position. The
X-ray source appears extended with a deconvolved size of $\sim
5^{\prime\prime}$ (the XMM point spread function is $\sim
6^{\prime\prime}$). The total X-ray flux is $F_X = 9.1 \times 10^{-18}
\rm W\,m^{-2}$, which would mean a luminosity of $L_X = 8.9 \times
10^{36} \rm W$ if the emission is associated with the cluster at
$z=1.28$ or $L_X = 1.2 \times 10^{37} \rm W$ at $z=1.45$. This would
be indicative of a cluster mass of $\sim 10^{14} \rm M_{\odot}$ for
one of the clusters, or $\sim 8 \times 10^{13} \rm M_{\odot}$ each if
both clusters contribute equally to the X-ray flux (Reiprich \&
B\"ohringer 2002). The signal-to-noise ratio is too low to allow us to
detect the 6.4 keV iron line and thus we cannot assign a secure
redshift to the X-ray source. However, enough photons are detected in
the source to conduct a crude hardness ratio (HR) spectral
analysis. The HR-values are HR2 $=(C_3 - C_2) / (C_2 + C_3) =
0.38\,\pm\,0.22 $ and HR3 $= (C_4 - C_3) / (C_3 + C_2) =
0.01\,\pm\,0.16$ where $C_2$ is the 0.5 - 2.0 keV count rate, $C_3$
the 2.0 - 4.5 keV count rate, and $C_4$ the 4.5 - 10.0 keV count rate
(Ueda et al. [in prep.]). These ratios mean the X-ray spectrum is much
harder than typical for a cluster, and is more characteristic for an
absorbed active galactic nucleus (e.g. Della Ceca et
al. 2004). However, we do not find evidence for a single obscured AGN
(e.g. a 24-$\mu$m or radio source) at the exact position of the X-ray
emission. This is a puzzling system which is at face value an extended
hard X-ray source. It requires an emission mechanism beyond any simple
addition of a single obscured, non-thermal AGN and thermal cluster
emission. Two such mechanisms that can cause the extended hard X-ray
emission are inverse Compton scattering (see for example Fabian et
al. 2006b) or confusion of more than one hard-spectrum X-ray source.

There is a significantly brighter X-ray source $\sim 40$ arcsec to the
southwest of CVB13 (X1, see Fig.~\ref{overlay} [{\it left}]). It is
uncertain if it is physically associated with the weaker extended
X-ray emission; however it is certainly associated with a distant
galaxy. A third X-ray source in the field (X3) is associated with an
AGN at a redshift coincident with the $z=1.28$ structure (see
Table~\ref{objects}). Table~\ref{objects} also shows the properties of
the eight brightest radio sources in the field shown in
Fig.~\ref{overlay} ({\it left}). There are radio sources associated
with both the $z=1.28$ and the $z=1.45$ clusters, but the bright radio
source RB4 closest to CVB13 is associated with a foreground galaxy at
$z \sim 0.5$. In Fig.~\ref{overlay} ({\it right}) we plot a zoomed-in
view of the CVB13 field and overplot the contours of the VLA
A-configuration map. This confirms four weak radio sources tentatively
seen in the B-configuration map (Fig.~\ref{overlay} [{\it left}]). One
of these objects (RA3) has spectroscopically been confirmed to be at
$z=1.45$. RA4 has similar colours and photometric redshift and is
therefore likely to be part of the same cluster. There is inconclusive
evidence that the radio source associated with this galaxy has the
`head-tail' structure characteristic of galaxies moving with respect
to an intracluster medium.

If the radio emission is solely due to synchrotron radiation from a
starburst, the SFR of a radio source with a 1.4-GHz flux density of
$100 \,\mu\rm Jy$ is $\sim 1400 \,\rm M_{\odot}\,yr^{-1}$ in stars
with masses of $> 5 \,\rm M_{\odot}$ (see Condon 1992). Assuming a
Salpeter IMF, this would mean a SFR($M > 0.1\,\rm M_{\odot})$ of $\sim
7000 \,\rm M_{\odot}\,yr^{-1}$. The SFR derived from the \oii~
emission of these objects is only $\sim 10 \,\rm M_{\odot}\,yr^{-1}$,
and although this estimate does not take extinction into account and
is therefore a lower limit, the sub-millimetre SCUBA HAlf Degree
Extragalactic Survey (SHADES, e.g. Coppin et al. 2006) covers
this region and does not identify any highly obscured starbursts. If,
on the other hand, all radio emission is ascribed to AGN activity we
would expect (using Simpson et al. 2006, Eq. 9) a total X-ray
luminosity of $\sim 10^{-16} \rm W\,m^{-2}$ in the 2 - 10 keV
range. The object X2 only has a flux in this band of $\sim 10^{-17}
\rm W\,m^{-2}$; therefore confusion between hard X-ray sources
associated with the radio sources is a possible explanation for this
seemingly extended hard X-ray source.

\section{Concluding Remarks}
We have obtained multi-object spectroscopy on a photometrically
selected cluster candidate at $z=1.4$. Instead of a single massive
cluster, we find three projected structures of galaxies at $z=1.40$,
$z=1.45$, and $z=1.48$, one of which (at $z=1.45$) is statistically
robust. We also serendipitously detect a robust structure at
$z=1.28$. Both the structure at $z=1.45$ and the one at $z=1.28$ have
masses $\sim 10^{14} \rm M_{\odot}$ and may therefore be termed poor
galaxy clusters.
 
The evolutionary status of both the $z=1.28$ and $z=1.45$ clusters is
uncertain. The cluster at $z=1.45$ has a significantly lower virial
mass than its mass derived from the total observed luminosity. It is
very likely that this structure is not yet virialised, which would
suppress the derived (three-dimensional) velocity dispersion of $\sim
200 \rm\,km\,s^{-1}$ . The velocity shear (half-width to zero
intensity) observed in CVB13\_21 - located centrally within the
cluster - is $320 \rm\,km\,s^{-1}$, which shows that the measured
velocity dispersion of the cluster is unlikely to represent the total
cluster mass. Note however that low-number statistics make an accurate
determination of the velocity dispersion unfeasible. The total
measured SFRs are $\sim 45$ and $\sim 60\,\rm M_{\odot}\,yr^{-1}$
respectively for the clusters at $z=1.28$ and $z=1.45$, which for a
mass of $\sim 1 \times 10^{14} \rm M_{\odot}$ within $4 \times 4
\times 4\,\rm Mpc^3$ (comoving) is a factor of 3 - 5 higher than
typical star-formation rates observed in the field at these redshifts
(e.g. Conolly et al. 1997)

The analysis of the cluster galaxies in this paper is based solely on
redshifts estimated from \oii~emission line features. Therefore the
overdensities we find at $z=1.28$ and $z=1.45$ only consist of
star-forming galaxies, whereas the candidate cluster found by VB06 was
selected with a photometric redshift algorithm which is sensitive to
red, passively-evolving galaxies. Of the 18 highest-priority targets
of sample 1, 10 were observed, and only six redshifts were obtained,
two of which were confirmed to be cluster members. The other galaxies
were all priority 3 targets of sample 1. We suspect there will be an
appreciable number of red galaxies in the cluster that we were unable
to confirm spectroscopically. More observations of the central cluster
galaxies, specifically for the cluster at $z=1.45$, may also produce a
higher velocity dispersion measurement. To thoroughly understand the
nature and masses of the structures, we need to be able to obtain
redshifts for this type of galaxy as well. Deep, near-infrared
spectroscopy, for example with the Fibre Multi Object Spectrograph
(FMOS, Dalton et al. 2006), will give a more complete sample and
therefore yield a more comprehensive result for high-redshift galaxy
clusters.

The extended X-ray luminosity at the position of these structures is
characteristic of a cluster with a mass of $\sim 10^{14} \rm
M_{\odot}$. However the X-ray spectrum is too hard to simply be caused
by thermal cluster gas emission. A non-thermal effect like inverse
Compton scattering could cause an X-ray detection with a spectral
shape reflecting the energy distribution of the underlying electron
population. The most likely explanation, however, is confusion caused
by more than one AGN contributing to the X-ray emission. To shed light on
the exact nature of this system, it will be interesting to study the
cluster gas via the Sunyaev-Zel'dovich effect and the dark-matter mass
via lensing.

The original mass estimate of CVB13 by VB06 of $3 \times 10^{14}\rm
M_{\odot}$ is likely to be overestimated due to the superposition of
the two lower-mass structures at $z=1.40$ and $z=1.48$. The single
structure at $z=1.28$ contains too few galaxies to be isolated by the
same selection algorithm. This result suggests that spectroscopic
follow-up is a vital element of photometric cluster surveys as the
latter are prone to projection effects and can seriously overestimate
cluster masses. To determine the impact of these effects, we calculate
the probability of finding three groups in projection within $\Delta z
= 0.1$ in a survey such as the one presented in VB06. Assuming a
Sheth-Tormen cluster mass function (Sheth \& Tormen 1999) we expect to
find 121 groups with masses of a few $10^{13} \rm M_{\odot}$ and 6
clusters with masses of $\sim 10^{14} \rm M_{\odot}$ within the
redshift range $1.25 \leq z \leq 1.55$ in a field of 0.5
deg$^2$. Neglecting spatial clustering of the groups, there are on
average two superpositions of three groups with masses $\sim 10^{13}
\rm M_{\odot}$ within a box of 2 x 2 Mpc$^2$ and $\Delta z= 0.1$. The
probability of finding a projection of two groups and one cluster of
$\sim 10^{14} \rm M_{\odot}$ (as for CVB13) is 45 per cent. Running the
cluster algorithm of VB06 on a simulated superposition of three groups
shows that in order to distinguish between massive clusters at this
redshift and projected groups with a redshift separation of $\Delta z
= 0.05$, the photometric redshift error needs to be as small as
$\frac{\sigma_z}{1+z} = 0.005$. The COMBO-17 survey (Wolf et
al. 2004), which determined photometric redshifts from photometry in
17 bands, obtained a best redshift error of $\frac{\sigma_z}{1+z} =
0.01$ for galaxies with $R < 21$ (c.f. $R \sim 24$ for CVB13
targets). Thus attaining the order of accuracy needed in photometric
redshifts to determine the cluster mass function at high redshifts
seems very challenging, and we caution against attempting this with
photometric techniques alone.

\section{Acknowledgments}
The authors are grateful to Lance Miller for sharing his cluster mass
function code. CvB would like to thank Oxford Astrophysics for
studentship funding and CvB, DB, and CS acknowledge funding from the
Science and Technology Facilities Council. The data presented herein
were obtained at the W.M. Keck Observatory, which is operated as a
scientific partnership among the California Institute of Technology,
the University of California and the National Aeronautics and Space
Administration. The Observatory was made possible by the generous
financial support of the W.M. Keck Foundation. The authors wish to
recognise and acknowledge the very significant cultural role and
reverence that the summit of Mauna Kea has always had within the
indigenous Hawaiian community.  We are most fortunate to have the
opportunity to conduct observations from this mountain.  The analysis
pipeline used to reduce the DEIMOS data was developed at UC Berkeley
with support from NSF grant AST-0071048.

\end{document}